\documentclass[showpacs,twocolumn,superscriptaddress,aps,prb]{revtex4-1}

\usepackage[FIGTOPCAP,raggedright,normalsize,tight]{subfigure}

\usepackage{graphicx}
\usepackage{dcolumn}
\usepackage{bm}
\usepackage{amsmath}
\usepackage{color}

\begin{document}

\title{Stable hydrogenated graphene edge types: Normal and reconstructed Klein edges}

\author{Philipp Wagner}
\affiliation{Institut des Mat\'eriaux Jean Rouxel (IMN), Universit\'e de Nantes, CNRS UMR 6502, 44322 Nantes, France}

\author{Viktoria V. Ivanovskaya}
\affiliation{Unit\'e Mixte de Physique CNRS-Thales, 91767 Palaiseau, and Universit\'e Paris-Sud, 91405 Orsay, France}

\author{Manuel Melle-Franco}
\affiliation{Departamento de Inform\'atica, Centro de Ci\^encias e Tecnologias da Computa\c{c}\~ao, Universidade do Minho, 4710-057 Braga, Portugal}

\author{Bernard Humbert}
\author{Jean-Joseph Adjizian}
\affiliation{Institut des Mat\'eriaux Jean Rouxel (IMN), Universit\'e de Nantes, CNRS UMR 6502, 44322 Nantes, France}

\author{Patrick R. Briddon}
\affiliation{School of Electrical, Electronic and Computer Engineering, University of Newcastle, Newcastle upon Tyne, NE 1 7RU, United Kingdom}

\author{Christopher P. Ewels}
\email{chris.ewels@cnrs-imn.fr}
\affiliation{Institut des Mat\'eriaux Jean Rouxel (IMN), Universit\'e de Nantes, CNRS UMR 6502, 44322 Nantes, France}

%\date{\today}

\begin{abstract}

Hydrogenated graphene edges are assumed to be either armchair, zigzag or a combination of the two. We show that the zigzag is not the most stable fully hydrogenated edge structure along the ${<}2\bar{1}\bar{1}0{>}$ direction. Instead hydrogenated Klein and reconstructed Klein based edges are found to be energetically more favourable, with stabilities approaching that of armchair edges. These new structures `unify' graphene edge topology, the most stable flat hydrogenated graphene edges always consisting of pairwise bonded C$_2$H$_4$ edge groups, irrespective the edge orientation. When edge rippling is included, CH$_3$ edge groups are most stable. These new fundamental hydrogen terminated edges have important implications for graphene edge imaging and spectroscopy, as well as mechanisms for graphene growth, nanotube cutting, and nanoribbon formation and behaviour.

\end{abstract}

%See also Phys. Rev. B 88, 094106, 2013


\maketitle

\begin{figure}[bp]
\centering
\subfigure[]{\includegraphics[width=0.52\linewidth]{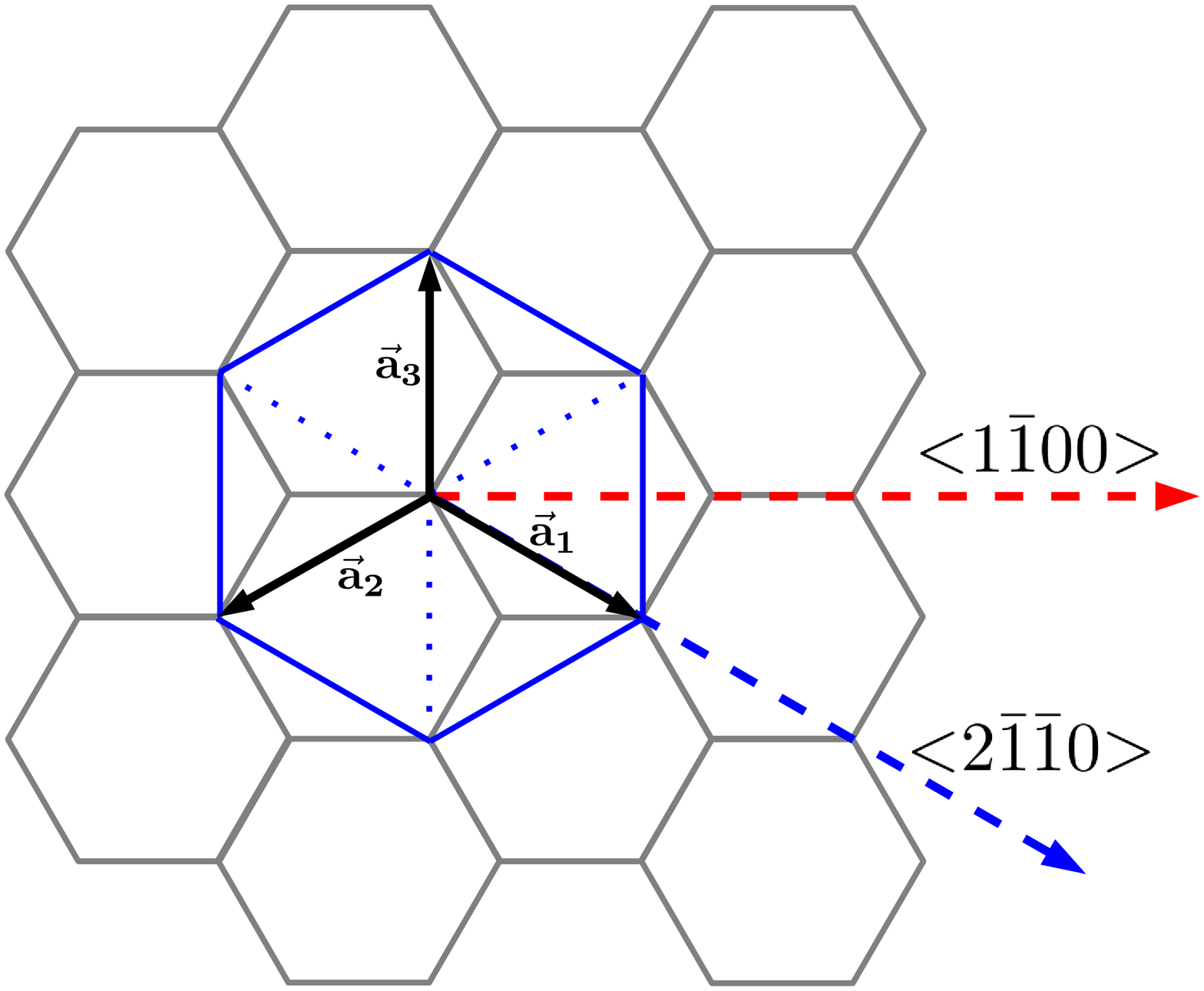}}
\subfigure[]{\includegraphics[width=0.46\linewidth]{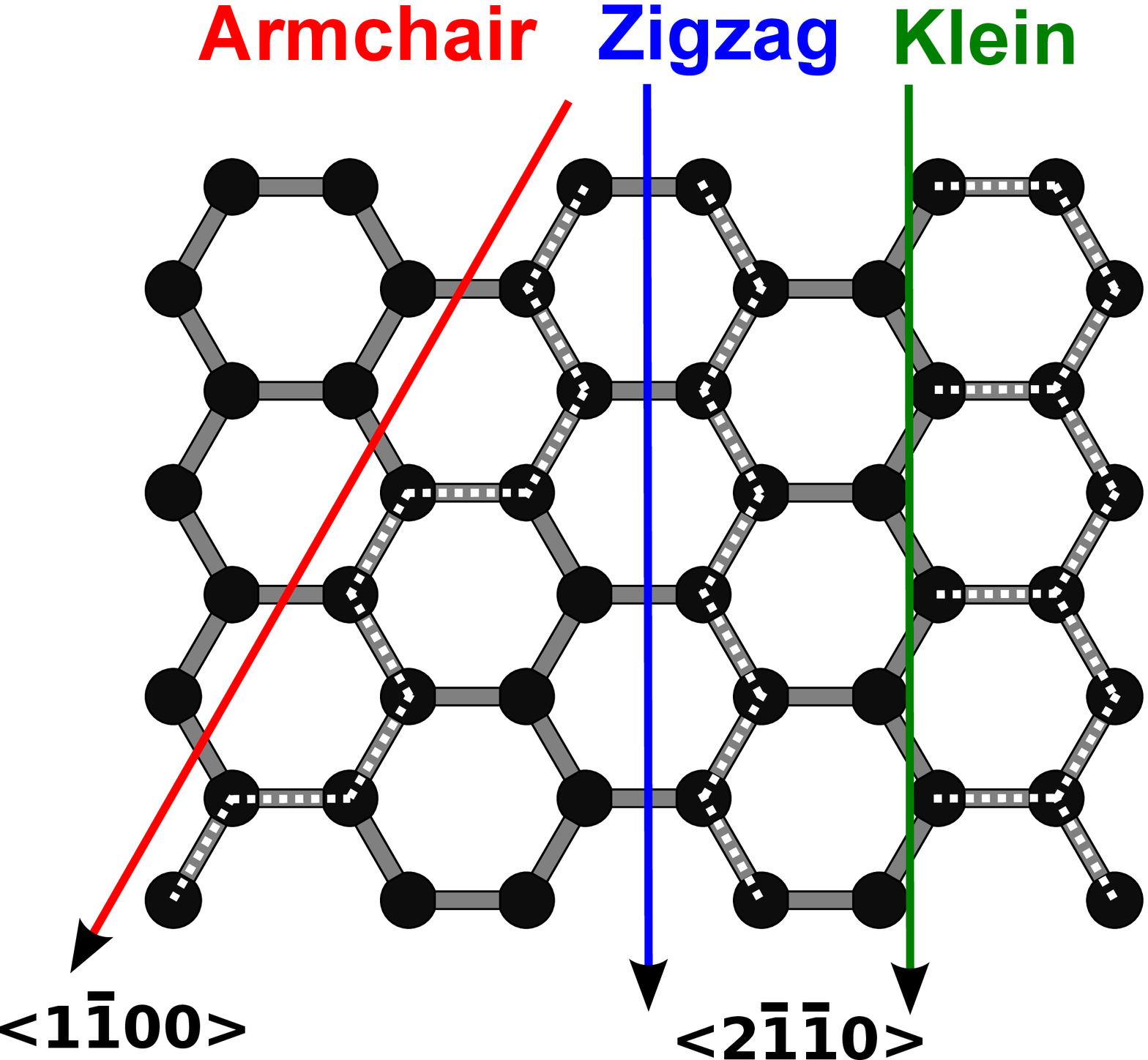}}
\caption{(a) Top view of the hexagonal system in the graphene honeycomb lattice plane, with highlighted characteristic directions ${<}1\bar{1}00{>}$ and ${<}2\bar{1}\bar{1}0{>}$. (b) Cutting a graphene sheet to create an armchair (red), zigzag (blue) or Klein edges (green). The resulting atomic edge structures are marked via dotted white lines to the right of the cutting axes.}
  \label{cut_graphene}
\end{figure}

Graphene edges have been studied intensively since the first interest in graphitic nanomaterials and graphene \cite{Nakada1996,Wakabayashi1999,Cancado2004,Geim2007,CervantesSodi2008,Wassmann2008,Koskinen2008,Girit2009,Jia2011}.
As recently shown, the precise edge termination of graphene nanoribbons and flakes has a significant effect on the material properties \cite{Ritter2009,Gan2010,Wagner2011a,Cocchi2011a}. The graphene honeycomb lattice can be cut along two primary directions, the ${<}1\bar{1}00{>}$ and the ${<}2\bar{1}\bar{1}0{>}$, creating so-called \textit{armchair} and \textit{zigzag}/\textit{Klein} edges respectively (see Fig.\ref{cut_graphene}). All intermediate orientations can be described in terms of alternating sections of these edge orientations \cite{Branicio2011}, in general called \textit{chiral} edges.\\
The armchair edge along the ${<}1\bar{1}00{>}$ direction is the most stable unreconstructed suspended graphene edge, both unterminated \cite{Koskinen2008,Ivanovskaya2011} and hydrogenated with each edge carbon bonded to two hydrogen atoms \cite{Wassmann2008,Koskinen2008}.\\
The ${<}2\bar{1}\bar{1}0{>}$ direction is more complex. Unlike the ${<}1\bar{1}00{>}$ direction, there exist two possibilities to terminate the graphene lattice parallel to the ${<}2\bar{1}\bar{1}0{>}$ direction (see Fig.\ref{cut_graphene} (b)), resulting in either the classic zigzag edge, or the Klein edge with only single neighbour edge carbon atoms \cite{Klein1994,Zobelli2012}. 
The largely ignored unterminated pristine Klein edge is highly unstable and can undergo pairwise reconstruction to a pentagon-terminated Klein edge \cite{Ivanovskaya2011}. This reconstructed edge nonetheless remains +0.4 eV/\AA \; less stable than unterminated armchair or (5-7) reconstructed zigzag edges \cite{Koskinen2008,Ivanovskaya2011}.\\ 
Further complexity arises for ${<}2\bar{1}\bar{1}0{>}$ edges when hydrogen terminated. 
Recent modelling has shown that the singly hydrogenated zigzag edge can be stabilised through adding a period three hydrogen pair-termination, the $z_{211}$ edge \cite{Wassmann2008}, with the resulting edge no longer magnetic \cite{Kunstmann2011}. The $z_{211}$ zigzag edge is to date the most stable ${<}2\bar{1}\bar{1}0{>}$ edge structure proposed in the literature.
This general focus on zigzag and armchair graphene edges means that there has been no discussion of hydrogenated Klein-based edge configurations so far.\\
We describe in this communication for the first time a study, based on density functional calculations, of hydrogen terminated Klein and reconstructed Klein edge configurations. Furthermore hydrogenated edge mixtures of reconstructed Klein and zigzag edge sections along the ${<}2\bar{1}\bar{1}0{>}$ direction are explored. These new edge structures are significantly more stable than hydrogen terminated zigzag edges.\\  

Spin polarised density functional calculations under the local density approximation were performed as implemented in the \textit{AIMPRO} code \cite{Briddon2000,Rayson2009,Briddon2011}.  The charge density is fitted to plane waves with an energy cut-off of 150 Ha (Ha: Hartree energy). Electronic level occupation was obtained using a Fermi occupation function with $kT = 0.04$ eV. Relativistic pseudo-potentials are generated using the Hartwingster-Goedecker-Hutter scheme \cite{Hartwigsen1998}, resulting in basis sets of 22 independent Gaussian functions for carbon and 12 for hydrogen. All calculations were performed using orthorhombic supercells, whose sizes have been checked and chosen to be sufficiently large to avoid interaction between neighbouring ribbons (vacuum distance between ribbons larger than $15$ \AA).  A fine k-point grid was chosen (armchair GNR: $(12 / N) \times 1 \times 1$ and zigzag GNR: $(18 / N) \times 1 \times 1$, where $N$ is the number of fundamental unit cells along the ribbon axis). Energies are converged to better than 10$^{-7}$~Ha. Atomic positions and lattice parameters were geometrically optimised until the maximum atomic position change in a given iteration dropped below 10$^{-5}$ a$_0$ (a$_0$: Bohr radius). Molecular dynamics calculations were performed with the self-consistent-charge density-functional tight-binding (SCC-DFTB) Hamiltonian \cite{Elstner1998} implemented in the DFTB+ code \cite{Aradi2007}. We run NVT Molecular Dynamics (MD) simulations using an Andersen thermostat with a re-selection probability of 0.2 and a time-step of 1 fs. \\

\begin{figure}
\centering
\begin{tabular*}{1.0\linewidth}{c c}
  \multicolumn{2}{c}{${<}2\bar{1}\bar{1}0{>}$ \hspace*{0.2cm}}\\
  \multicolumn{2}{c}{$\overbrace{\hspace*{8.1cm}}$ \hspace*{0.2cm}}\\
 (a) &  (b) \\
  Rec. Klein ($rk_{22}$) & Rec. Klein + Zigzag ($rk_{22}+z_{2}$) \\ 
  $E_{edge}=-0.030$ eV/\AA \; & $E_{edge}=-0.107$ eV/\AA \\
  \\
  \includegraphics[width=0.40\linewidth]{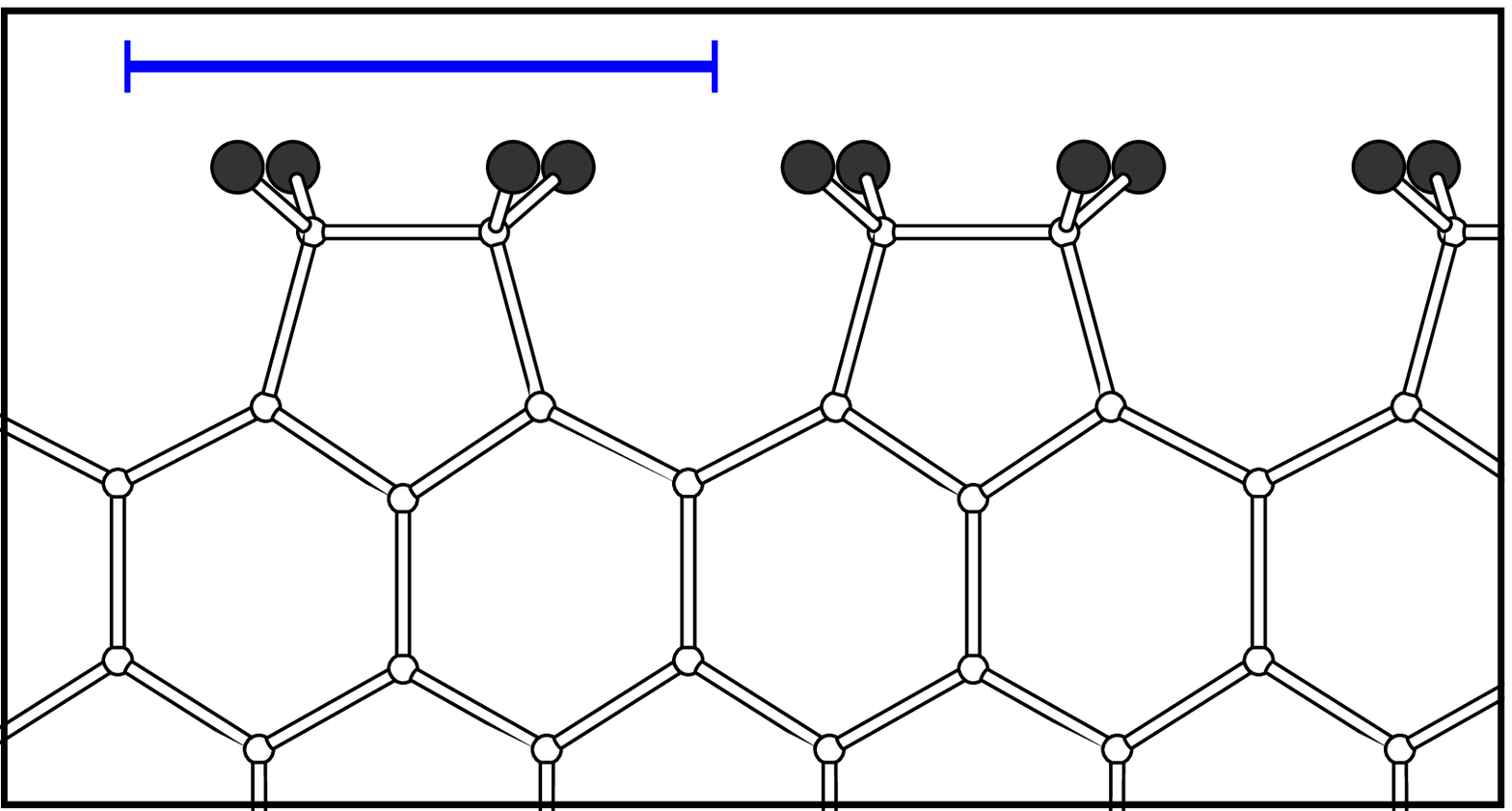} & \includegraphics[width=0.40\linewidth]{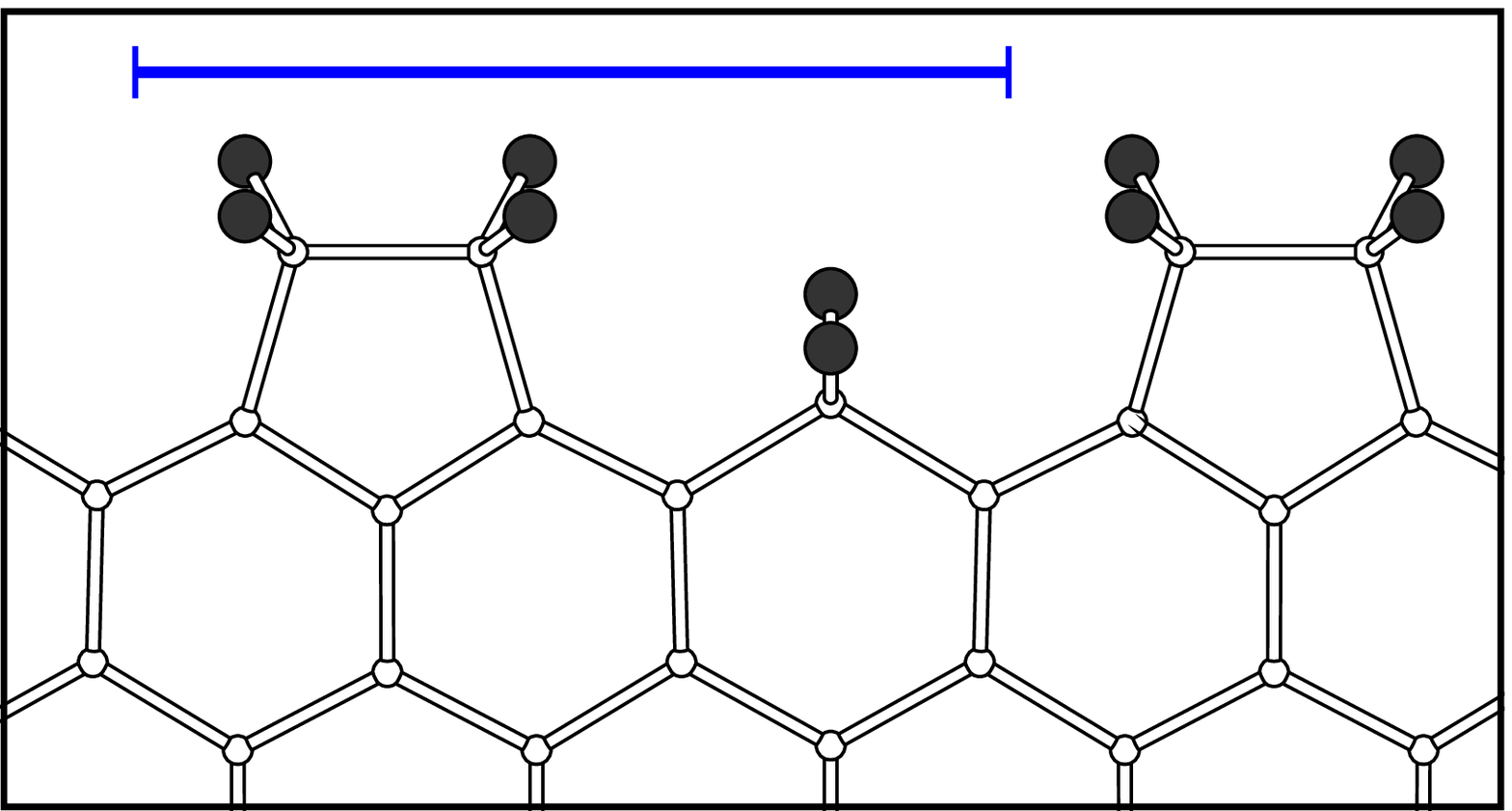} \\
   \includegraphics[angle=-90,width=0.45\linewidth]{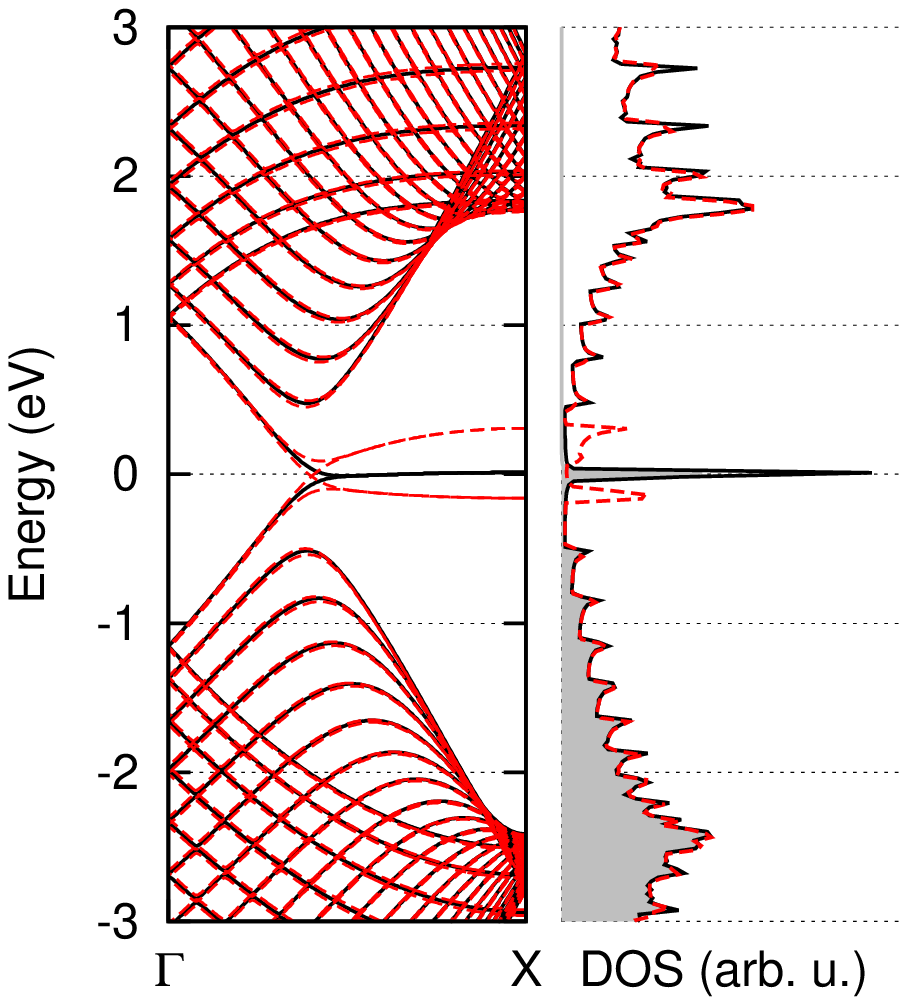} & \includegraphics[angle=-90,width=0.45\linewidth]{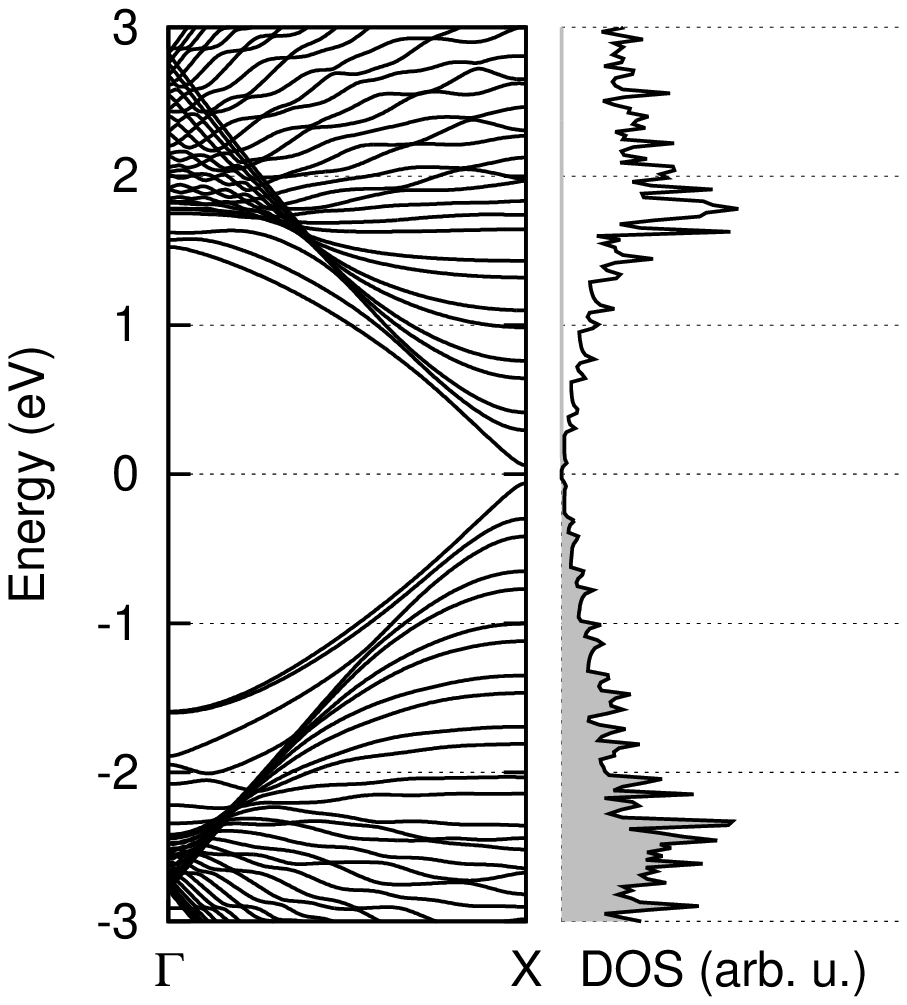} \\ 
  \\ 
   \hline
   \\
     ${<}1\bar{1}00{>}$ & ${<}2\bar{1}\bar{1}0{>}$  \\
  $\overbrace{\hspace*{4.0cm}}$ & $\overbrace{\hspace*{4.0cm}}$  \\
  (c) &  (d) \\  
  Armchair ($a_{22}$) & Zigzag ($z_{211}$) \\ 
  $E_{edge}=-0.186$ eV/\AA & $E_{edge}=-0.016$ eV/\AA \\ 
  \\
  \includegraphics[width=0.40\linewidth]{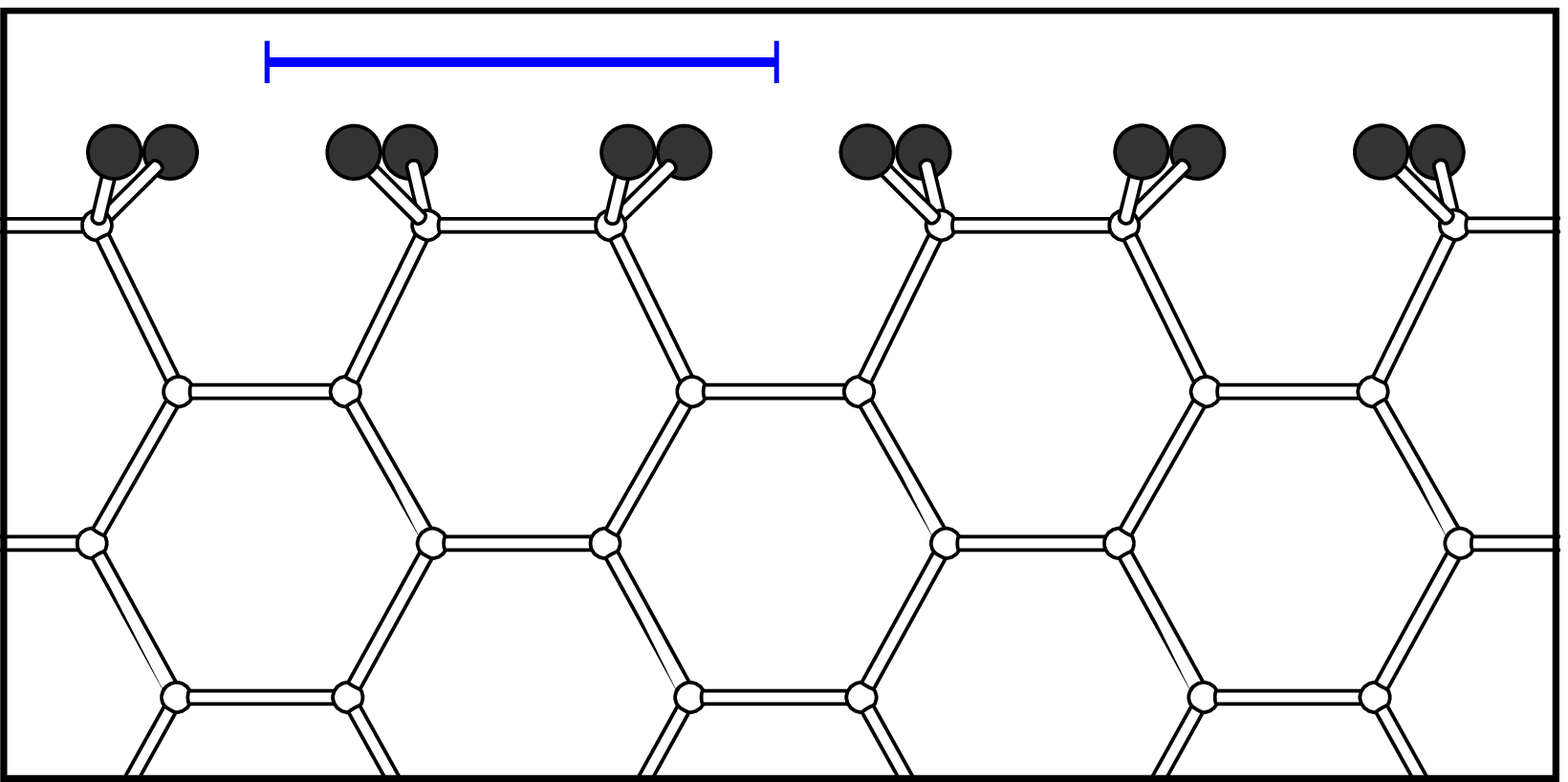} & \includegraphics[width=0.40\linewidth]{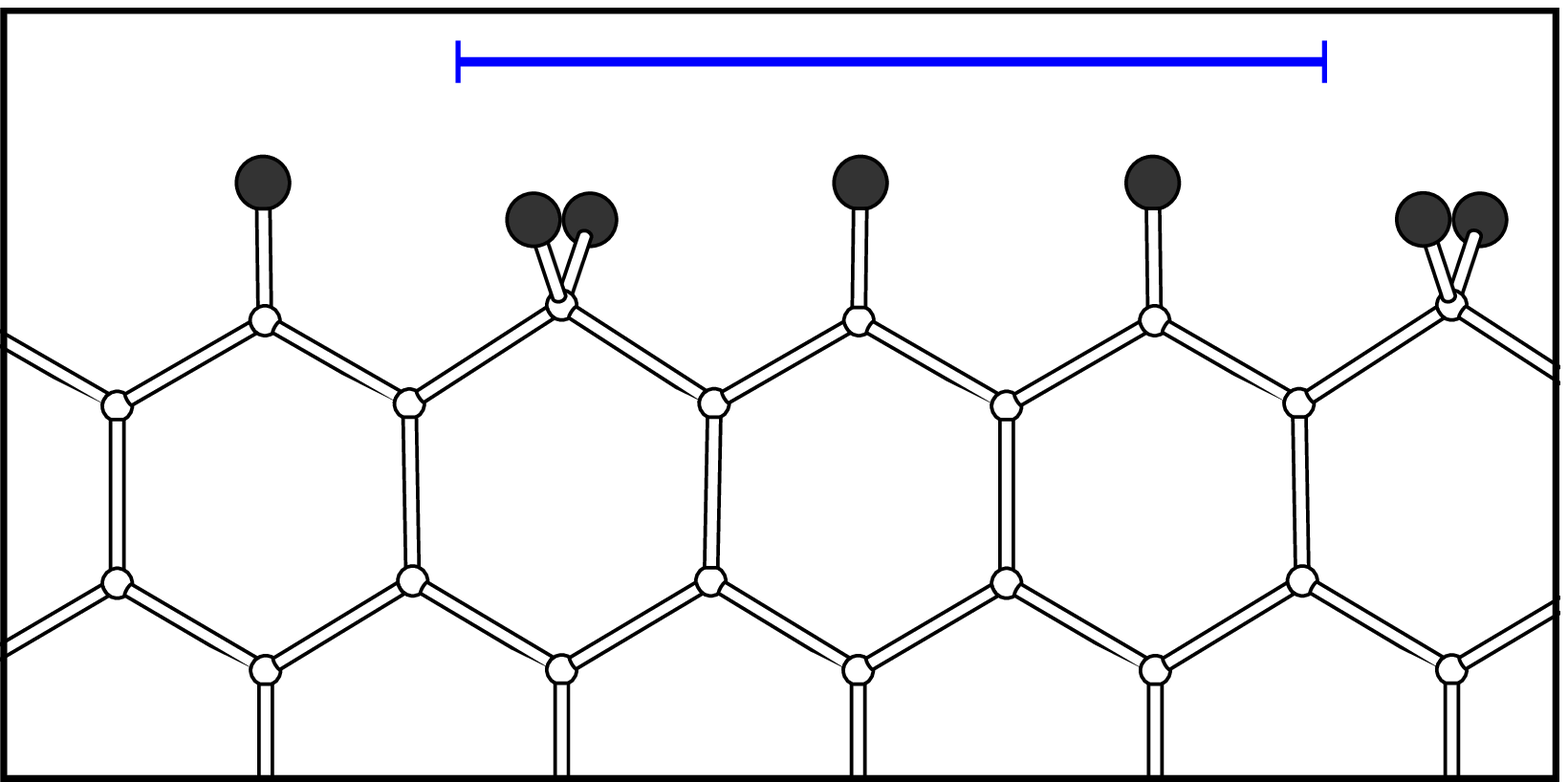} \\
   \includegraphics[angle=-90,width=0.45\linewidth]{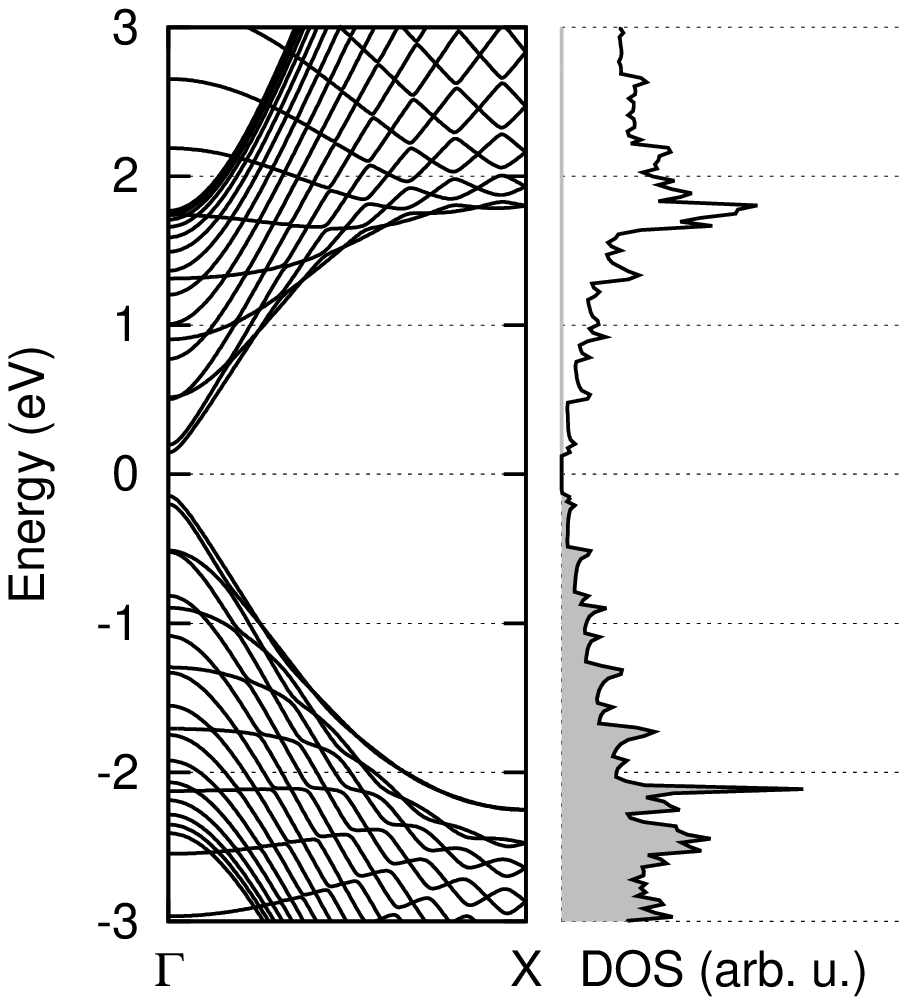} & \includegraphics[angle=-90,width=0.45\linewidth]{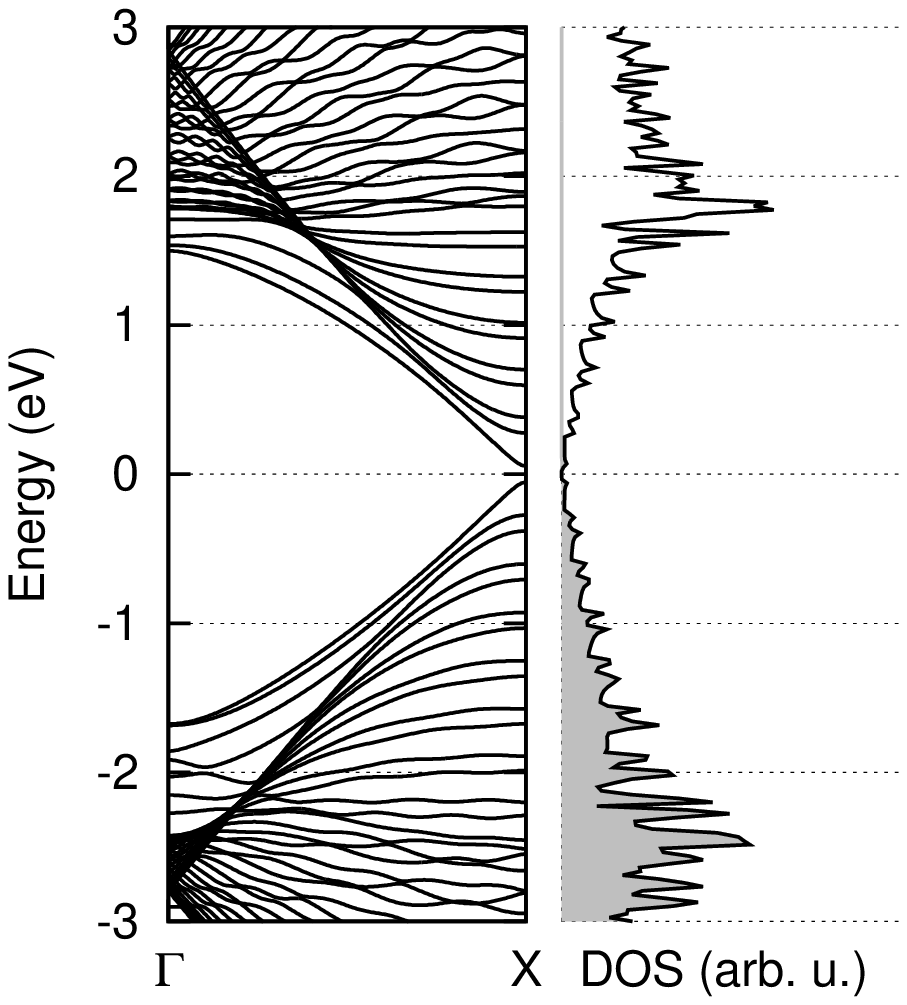} \\
 \end{tabular*}
\caption{Structures and edge formation energies ($E_{edge}$) of hydrogenated flat reconstructed Klein graphene edges, (a) $rk_{22}$ and (b)  $rk_{22}+z_{2}$, with associated band structure and density of states (DOS), Fermi level at 0 eV. The lower panel gives for reference the most stable hydrogenated (c) armchair $a_{22}$ and (d) zigzag $z_{211}$ edges with the corresponding band structures. Repeating segments are marked with a blue bar.  $rk_{22}$ is  $\approx$21 meV/unit cell more stable in a magnetic (red) than non-magnetic configuration (black). All other edges are most stable in the non-magnetic state. (H atoms are black spheres, C atoms are white circles.)}
\label{rk}
\end{figure}

\begin{figure}
\centering
\begin{tabular*}{1.0\linewidth}{c c}
  \multicolumn{2}{c}{${<}2\bar{1}\bar{1}0{>}$ \hspace*{0.2cm}}  \\
  \multicolumn{2}{c}{$\overbrace{\hspace*{8.1cm}}$ \hspace*{0.2cm}}
  \\
  (a) &  (b) \\
  Klein ($k_{33}^{ud}$) & Klein + Zigzag ($k_{33}^{ud}+z_2$) \\ 
  $E_{edge}=-0.150$ eV/\AA \;  & $E_{edge}=-0.191$ eV/\AA \\ 
  \\ 
  \includegraphics[width=0.40\linewidth]{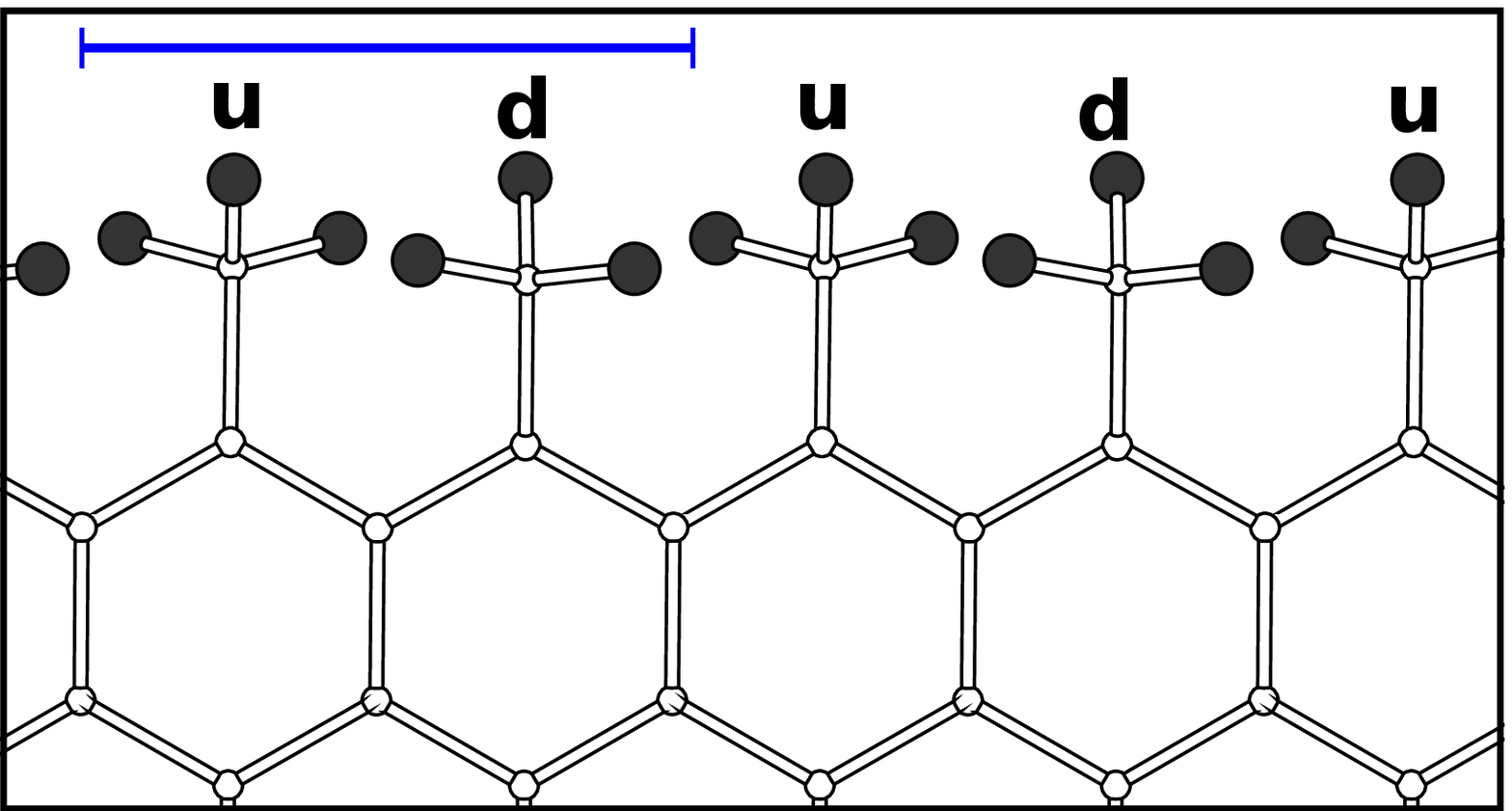} & \includegraphics[width=0.40\linewidth]{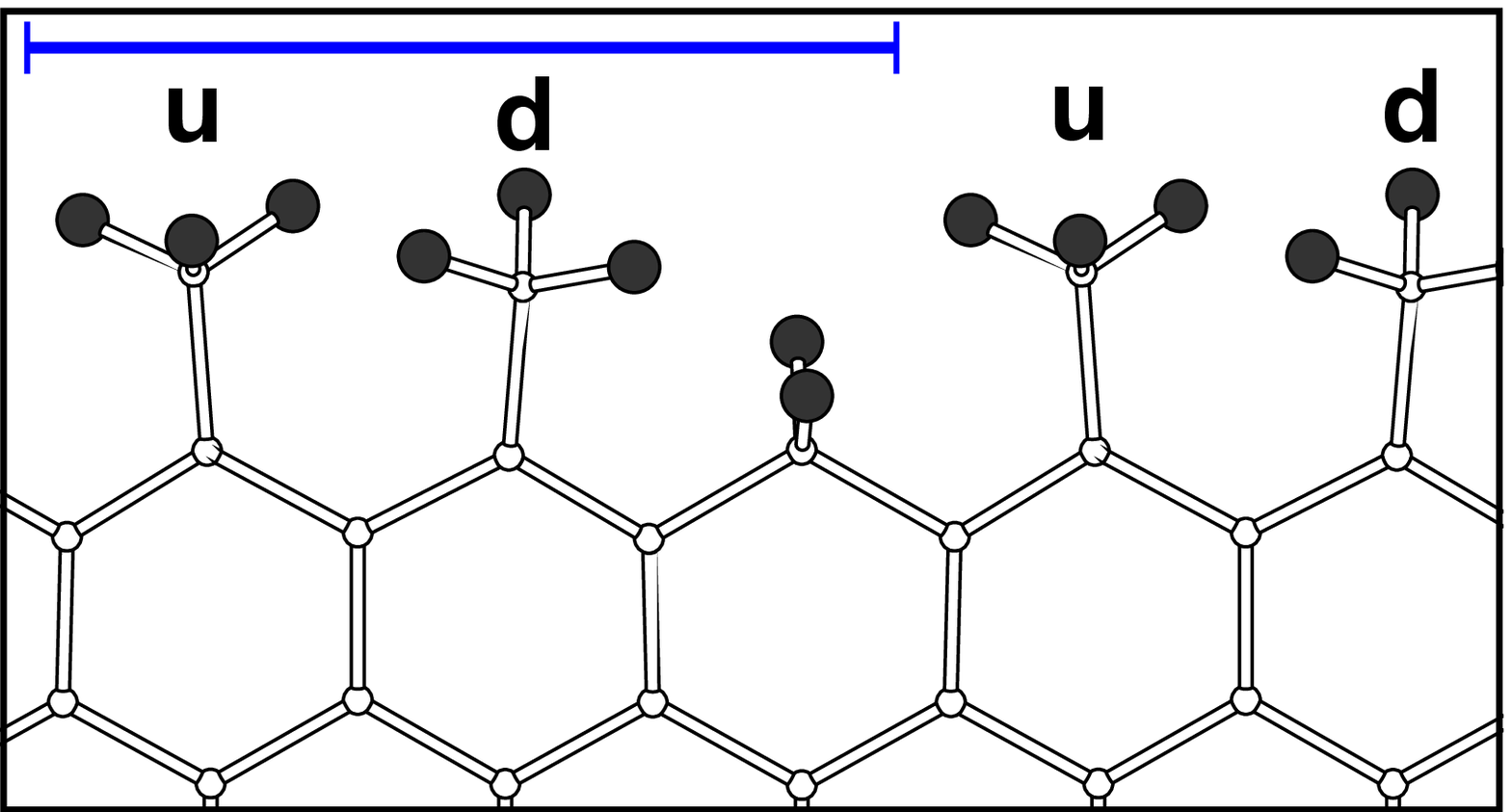} \\
  \includegraphics[angle=-90,width=0.45\linewidth]{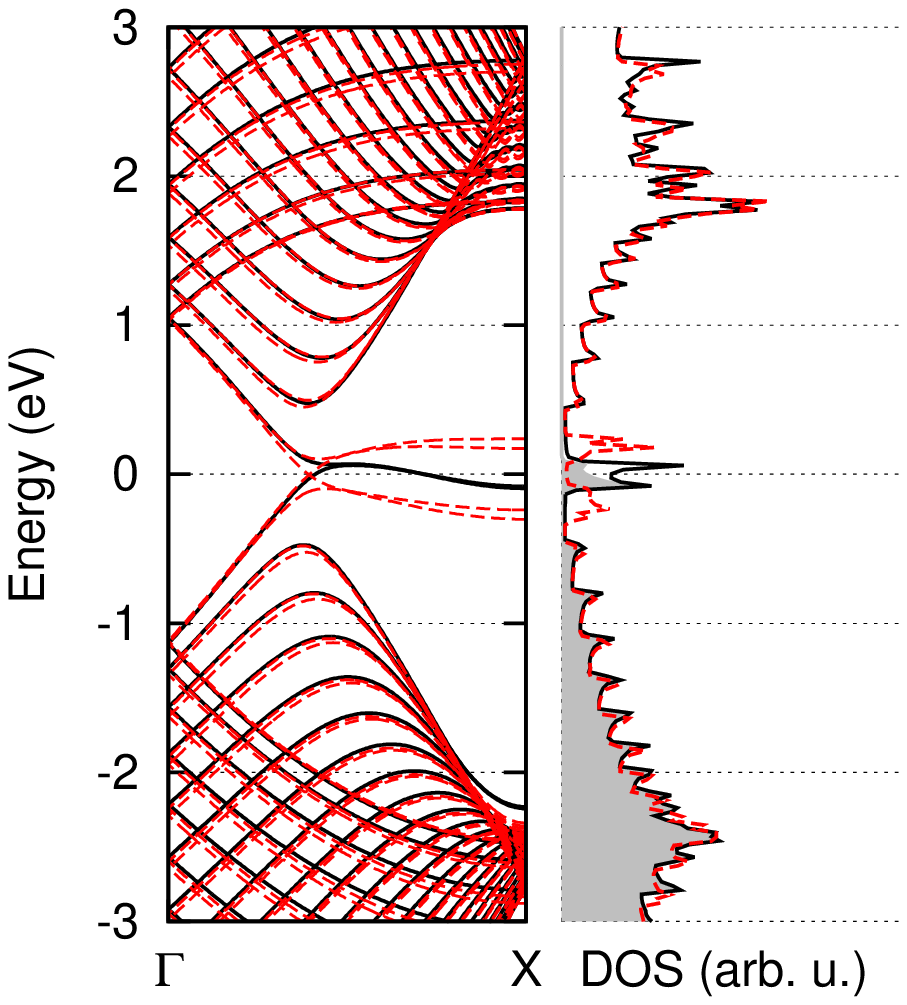} & \includegraphics[angle=-90,width=0.45\linewidth]{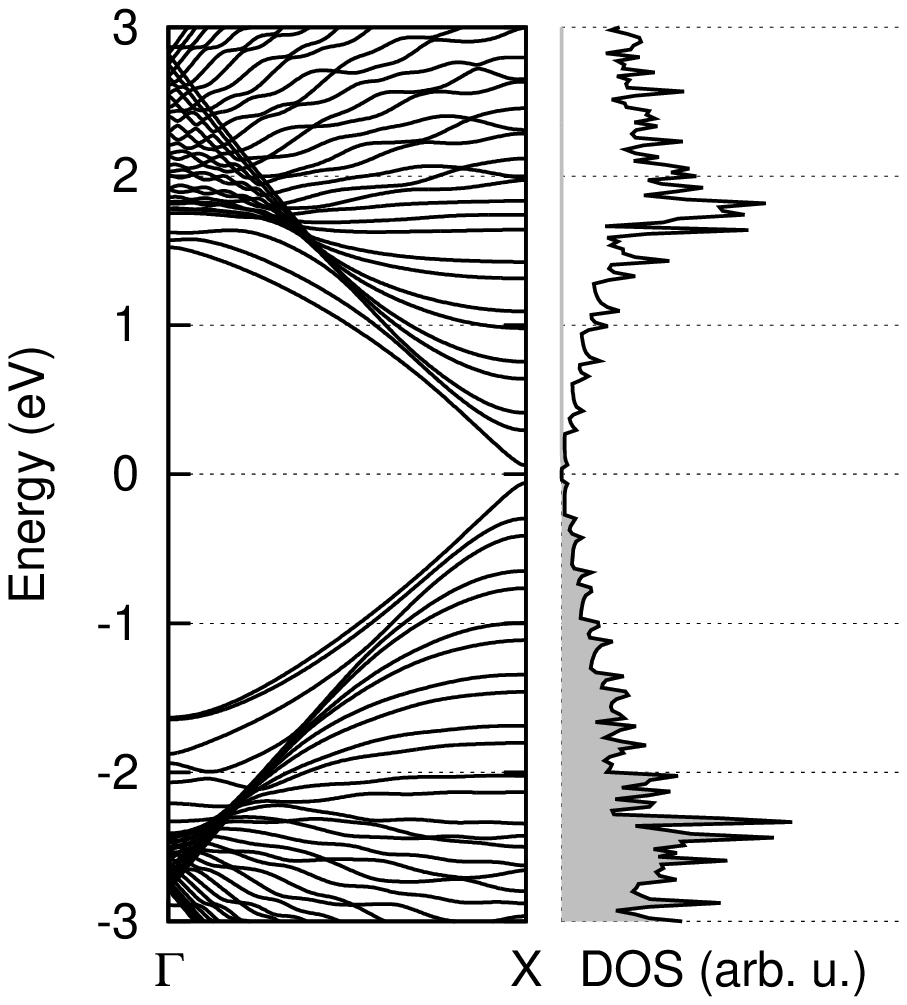} \\ 
 \end{tabular*}
\caption{Structure and formation energy $E_{edge}$ of the rippled fully hydrogenated (a) Klein $k_{33}^{ud}$ and (b) mixed Klein--zigzag $k_{33}^{ud}+z_2$ edge, with associated band structures and densities of states (DOS). $k_{33}^{ud}$ is  $\approx$65 meV/unit cell more stable in a magnetic (red) than non-magnetic configuration (black). The $k_{22}^{ud}+z_{2}$ edge is most stable when non-magnetic.}
\label{klein}
\end{figure}

Graphene nanoribbons (GNRs) of width $\sim$ 50 \AA \; have been used to model decoupled hydrogen terminated graphene edges. Ribbon segments in one unit cell consist of $n_C$ carbon atoms and $n_H$ hydrogen atoms. The edge formation energy $E_{edge}$ \cite{Wassmann2008,Ivanovskaya2011} is then calculated using:  
\begin{equation}
E_{edge} = \frac{E_{ribbon}-n_C \cdot E_C -n_H \cdot \frac{E_{H_2}}{2}}{2L} \; .
\label{formenergyH}
\end{equation}
Here, $E_{ribbon}$ is the total internal energy of the nanoribbon segment in a unit cell. $L$ is the length along the ribbon axis of the repeated nanoribbon segment (schematically indicated with blue bars in Fig.\ref{rk} and \ref{klein}), with two similar opposed graphene hydrogenated edge configurations. $E_C$ is the energy of a carbon atom in a perfect graphene sheet, $E_{H_2}$ gives the total internal energy of an isolated H$_2$ molecule.\\
Although several edge formation energies are found to be negative, this does not necessarily indicate the system favours edge formation over the bulk state under experimental conditions. Full free energy differences must be considered to address this, including entropy change and chemical potential of the edge components. Even if edge formation is thermodynamically favoured, reaction barriers may be prohibitive. Edge formation energies for armchair and zigzag edges are comparable with those in the literature \cite{Wassmann2008,Kunstmann2011,Koskinen2008,Lu2009}, with small quantitative differences due to choice of exchange correlation functional (for more details see Supplementary Materials \cite{supmat}).\\
A nomenclature to differentiate the various hydrogenated edge configurations is used similar to Wassmann et. al. \cite{Wassmann2008}, extended by the Klein and the reconstructed Klein edge ($a$: armchair, $z$: zigzag, $k$: Klein and $rk$: reconstructed Klein). Subscripts indicate the number of hydrogen atoms bonded to every edge carbon atom along the periodic edge segment. The new superscripts $u$ (``up'') and $d$ (``down'') indicate, where needed, out-of-plane edge deformations (edge rippling) \cite{Wagner2011a}.\\

The edge configurations $rk_{11}$ (+0.302 eV/\AA), $rk_{21}$ (+0.134 eV/\AA) and $rk_{22}+k_2$ (+0.034 eV/\AA) are all unstable, similar to the singly hydrogenated zigzag edge $z_1$ (+0.105 eV/\AA) (see also Supplementary Materials \cite{supmat}).
However, the hydrogenated reconstructed Klein edge $rk_{22}$ is energetically stable with an edge formation energy of $-0.030$ eV/\AA \; (see Fig.\ref{rk} (a)).  Notably the new hydrogenated $rk_{22}$ edge configuration is more than twice as stable as the most stable hydrogen terminated zigzag edge ($z_{211}$, $-0.016$ eV/\AA, Fig.\ref{rk} (d)), unequivocally proving that the most stable hydrogenated ${<}2\bar{1}\bar{1}0{>}$ oriented edges are not zigzag, but reconstructed Klein based.\\
The ground state for the reconstructed Klein edge $rk_{22}$ is magnetic (0.120 $\mu_B$/\AA), similar to the $z_1$ edge (0.128 $\mu_B$/\AA). \textit{Intra}-edge states couple ferromagnetically, while \textit{inter}-edge coupling \cite{Lee2005,Son2006a,Lee2009a,Kunstmann2011} is excluded here due to the large ribbon width. 
Magnetic zigzag edge states vanish with local defects or variations in hydrogen density (e.g. the $z_{211}$ edge \cite{Kunstmann2011}).  The perfect $rk_{22}$ edge should  protect the magnetic behaviour. We note that a stable magnetic (edge) state is an important property for graphene use in spintronics \cite{Yazyev2010}.\\

Just as the $z_1$ zigzag edge can be stabilised by periodically inserting double hydrogenation ($z_{211}$), so can the $rk_{22}$ edge via periodic insertion of a Klein edge vacancy, {\it i.e.} a double hydrogenated zigzag edge site $z_2$.  The resulting $rk_{22}+z_{2}$ edge has an edge formation energy of -0.107 eV/\AA \; (Fig.\ref{rk} (b)), approaching that of fully hydrogenated armchair edges with -0.186 eV/\AA \; (Fig.\ref{rk} (c)).  As for the $z_{211}$ edge this increased stability comes through the removal of the edge states around the Fermi level, resulting in a non-magnetic edge.\\

As expected, most hydrogenated unreconstructed Klein edges are thermodynamically unstable ($k_1$ (+1.276 eV/\AA), $k_2$ (+0.476 eV/\AA), $k_{32}$ (+0.052 eV/\AA) and $k_{332}^{ud}$ (+0.490 eV/\AA), see also Supplementary Materials \cite{supmat}). However the fully hydrogenated Klein edge $k_{33}^{ud}$ is very stable (-0.150 eV/\AA, see Fig.\ref{klein}).  Here the Klein edge carbon atoms are methylated giving CH$_3$, decoupling them from the sp$^2$ bonded graphene $\pi$-system. This stability is only possible  with periodic out-of-plane displacement of the edge methyl groups, relieving strain induced by inter-methyl steric hindrance \cite{Wagner2011a,Chia2012}.\\ 
As for the zigzag and reconstructed Klein edges, the $k_{33}^{ud}$ can be further stabilised through periodically inserting a Klein vacancy $k_{33}^{ud}+z_2$ (-0.191 eV/\AA), as shown in Fig.\ref{klein} (b). This results in a non-magnetic configuration with a slightly reduced out-of-plane rippling amplitude.  All three edge types ($z_{211}$, $k_{22}^{ud}+z_{2}$ and $k_{33}^{ud}+z_2$) then show periodic $sp^2-sp^2-sp^3$ bonding along their zigzag backbone.
We note that on metal surfaces, rippled fully hydrogenated Klein edges are likely to become less stable.  Dehydrogenation is also facilitated \cite{Celebi2013,Treier2011,Zhang2011b}, and dehydrogenated CH$_3$ edge groups could reconstruct to $rk_{22}$ edges.\\

The calculations presented thus far consider perfect vacuum conditions around free standing graphene edges.  A legitimate question arises, namely, what could be expected in experiments?
In order to consider a molecular hydrogen gas atmosphere around the graphene edge, the calculated total edge formation energy $E_{edge}$ can be compared to the hydrogen chemical potential $\mu_{H_2}$,
resulting in the relative edge stability
\begin{equation} 
G_{H_2} = E_{edge} - \rho_H \cdot \mu_{H_2}/2 \; . 
\label{edge_stability}
\end{equation}
Here, $\rho_H= \frac{n_H}{2L}$ gives the hydrogen edge density, with $n_H$ the number of
hydrogen atoms attached to a graphene ribbon segment of length $L$. The hydrogen chemical potential $\mu_{H_2}$ depends on the pressure and temperature of the system \cite{Wassmann2008}. As an indication, at ambient conditions the chemical potential $\mu_{H_2}$ at 300 K and partial $H_2$ pressure in air of $P_{H_2} \approx 5 \cdot 10^{-4}$ mbar gives $\mu_{H_2} \approx -0.4$ eV \cite{Chase1998}. For both, decreasing $P_{H_2}$ and increasing temperatures, the chemical potential decreases.\\
\begin{figure}
\includegraphics[width=1.0\linewidth]{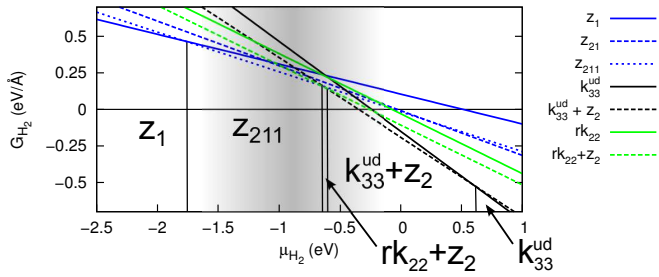}
\caption{Edge stability at different hydrogen chemical potentials $\mu_{H_2}$ for hydrogenated edges along the ${<}2\bar{1}\bar{1}0{>}$ direction. $G_{H_2} < 0$ indicates graphene instability due to hydrogen ``unzipping''. The gray region indicates typical substrate catalysed (ethylene) CVD growth conditions.}
\label{stability}
\end{figure}

\begin{figure}
\centering
\subfigure[]{\includegraphics[width=0.45\linewidth]{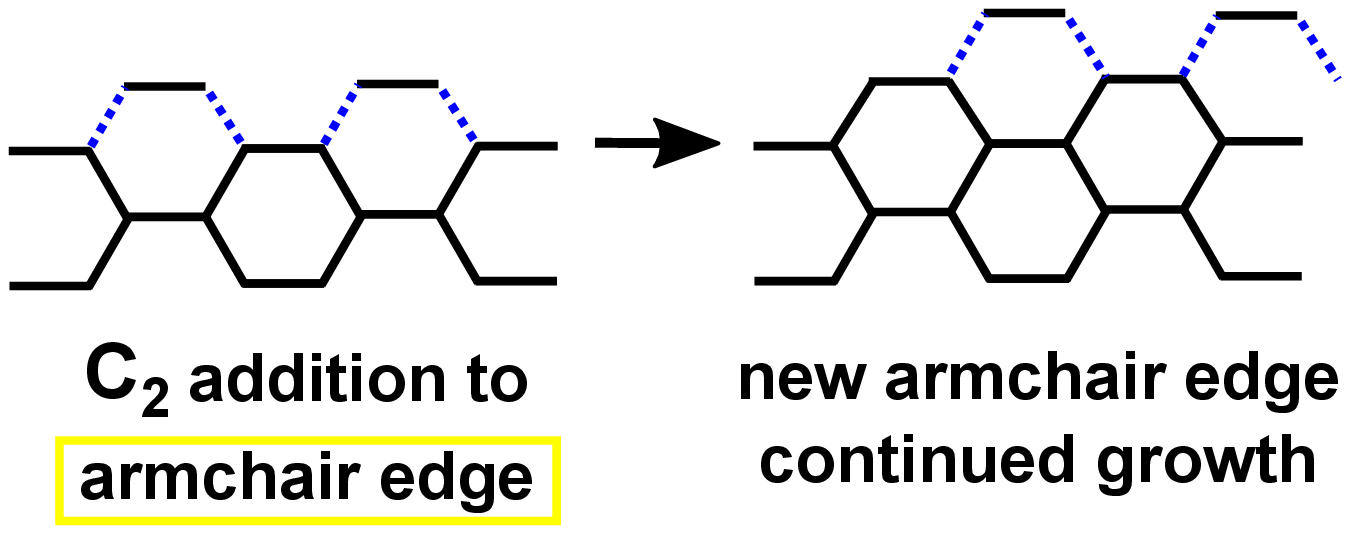}} \hspace*{0.1cm} \vline \hspace*{0.1cm}
\subfigure[]{\includegraphics[width=0.45\linewidth]{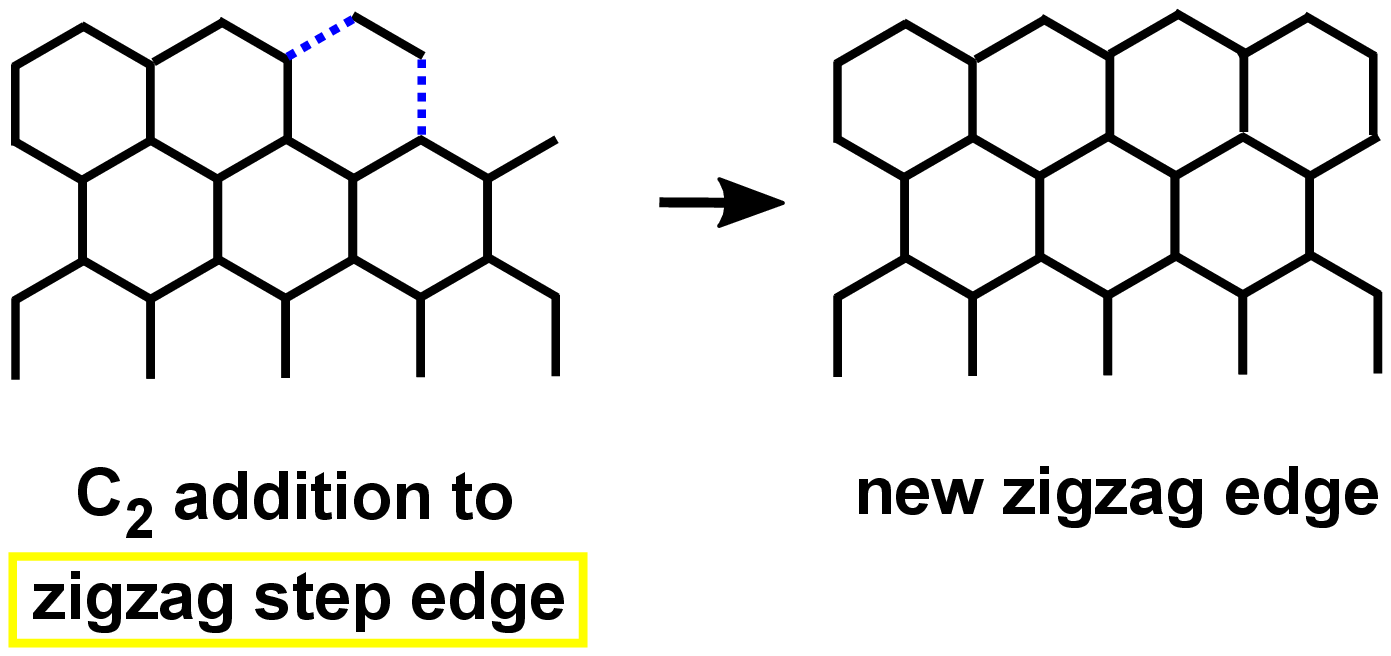}} \\ \vspace*{0.2cm}
\subfigure[]{\includegraphics[width=0.99\linewidth]{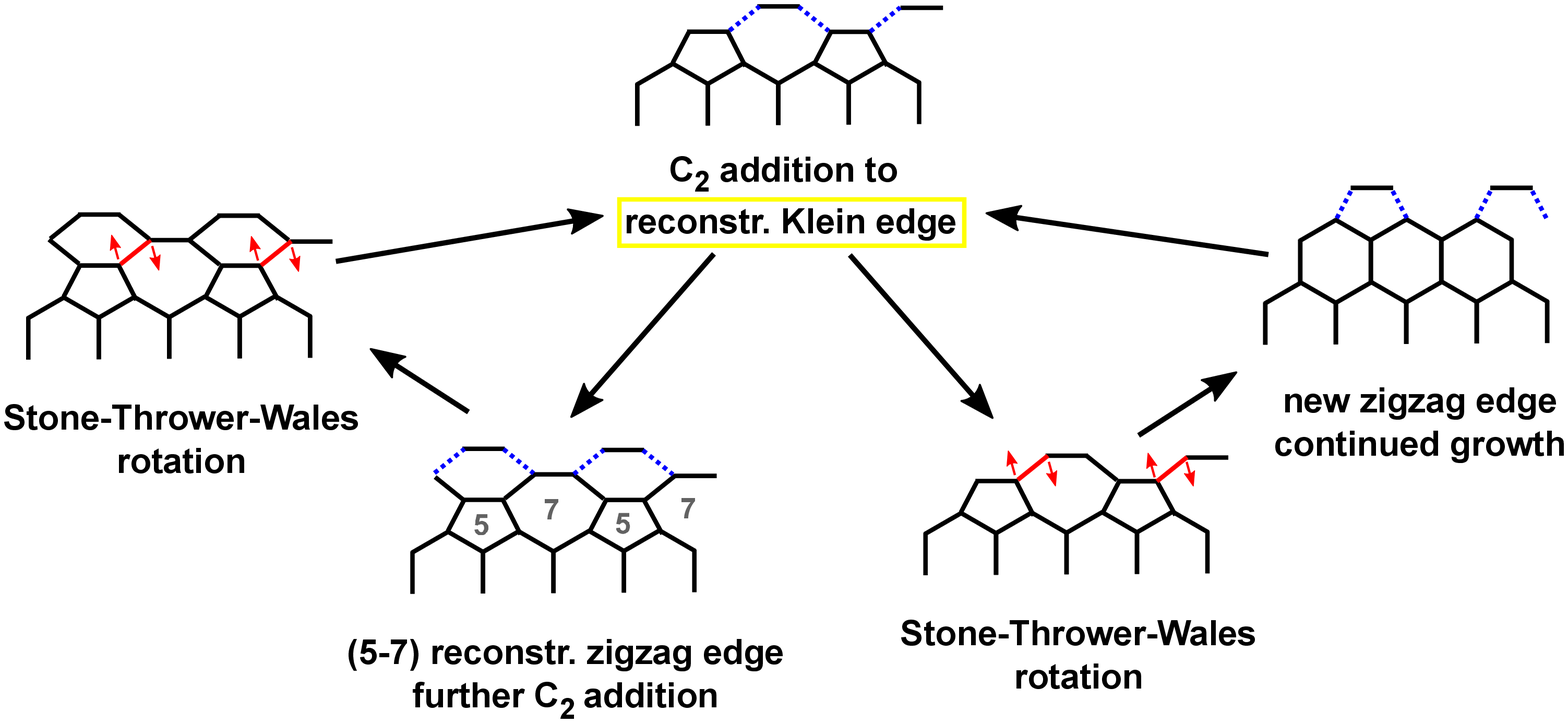}}
 \caption{Graphene growth models: (a) armchair edge, (b) zigzag step edge and (c) reconstructed Klein edge, with carbon dimers (C$_2$) as fundamental building blocks, similar to C$_2$H$_4$ (ethylene) precursors used for CVD graphene growth. Carbon dimer bonding to the graphene edges with dehydrogenation of the edge atoms marked with dotted blue lines, Stone-Thrower-Wales rotations are labelled red.}
\label{growth_model}
\end{figure} 
In Fig.\ref{stability}, $G_{H_2}$ is calculated over a realistic range of $\mu_{H_2}$, for the primary hydrogenated ${<}2\bar{1}\bar{1}0{>}$ edges, identifying regimes of different stable edge types. For low hydrogen chemical potentials the zigzag hydrogenated edges dominate. However, for higher pressures and/or moderate growth temperatures the new $rk_{22}+z_2$ and $k_{33}^{ud}+z_2$ configurations become favourable, starting around $\mu_{H_2} \geq -0.7$ eV. Up to $\mu_{H_2}=-0.4$ eV several different hydrogenated edge configurations are very close in energy, thermodynamically dominated by the new most stable identified Klein and reconstructed Klein configurations. Such low temperature and higher pressure conditions ($\mu_{H_2} \approx -1.2$ to $-0.5$ eV) are currently of interest for cost efficient CVD graphene growth on metal surfaces  \cite{Li2009a,Gao2010a,Celebi2011,Murdock2013}.  Given the close proximity in stability of these edge structures under typical CVD growth conditions, many of them may occur during growth. Possible growth mechanisms based on carbon dimers (C$_2$) as fundamental building blocks are proposed in Fig.\ref{growth_model}. Inclusion of reconstructed Klein edges opens the door to alternative reaction pathways including (5-7) reconstructed edges and Stone-Thrower-Wales bond rotations (e.g. Fig.\ref{growth_model} (c)).  These could provide an explanation for observed kinetic growth barriers on metal surfaces tentatively associated with graphene lattice construction \cite{Celebi2011}.\\

\begin{figure}
\centering
\includegraphics[width=1.0\linewidth]{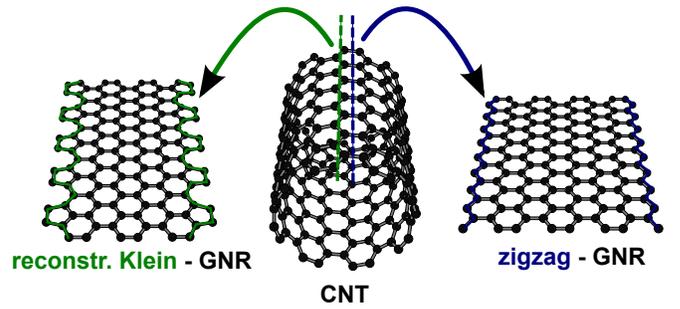}
  \caption{Schematic of cutting/unzipping a carbon nanotube (CNT) along the ${<}2\bar{1}\bar{1}0{>}$ direction to form either zigzag or reconstructed Klein graphene nanoribbons (GNRs). Hydrogen atoms not included for clarity.}
  \label{cut_CNT}
\end{figure}

\begin{figure}
\centering
 \subfigure[]{\includegraphics[width=0.49\linewidth]{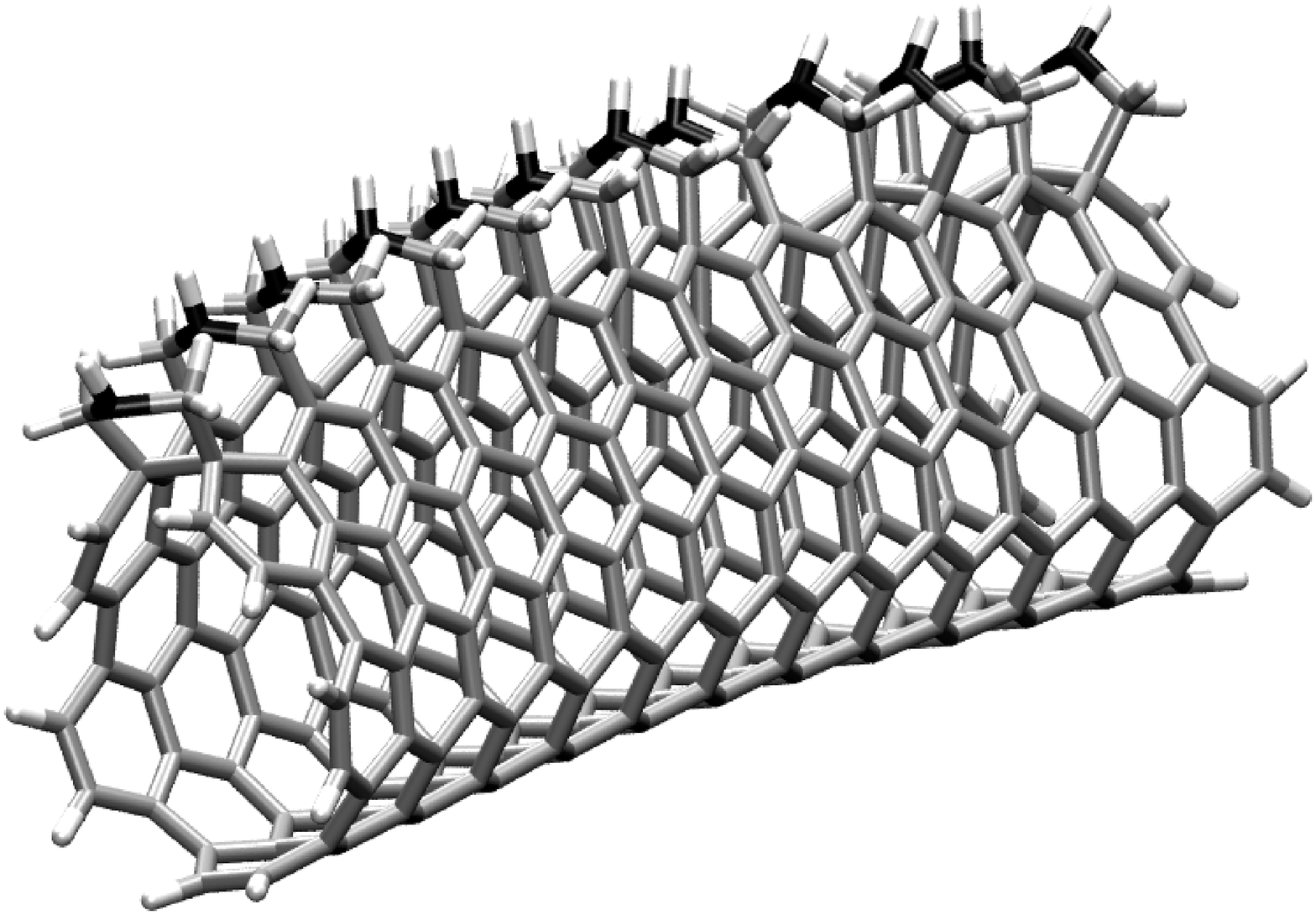}} \hspace*{0.3cm}
  \subfigure[]{\includegraphics[width=0.35\linewidth]{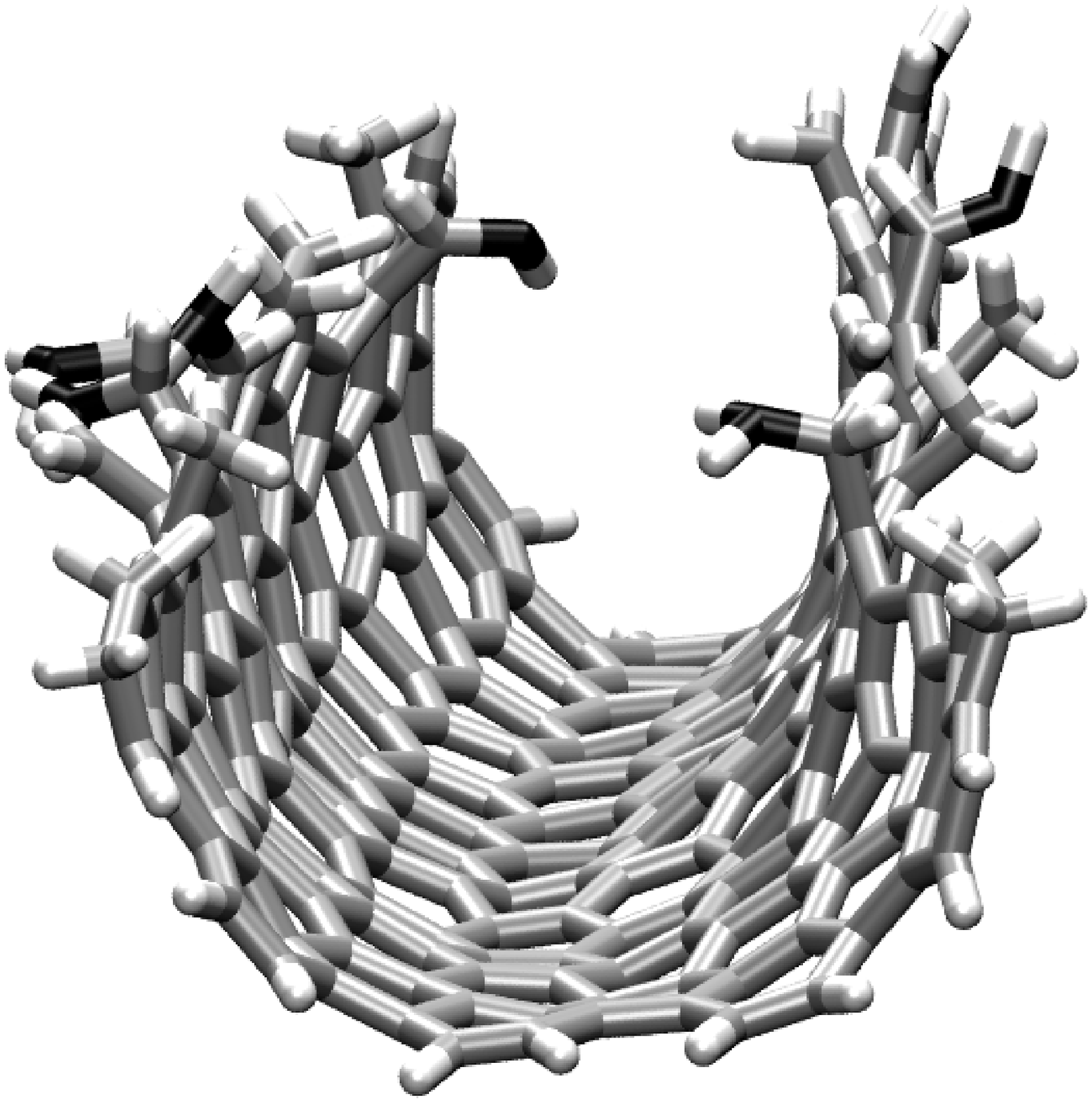}} \\
  \subfigure[]{\includegraphics[width=0.65\linewidth]{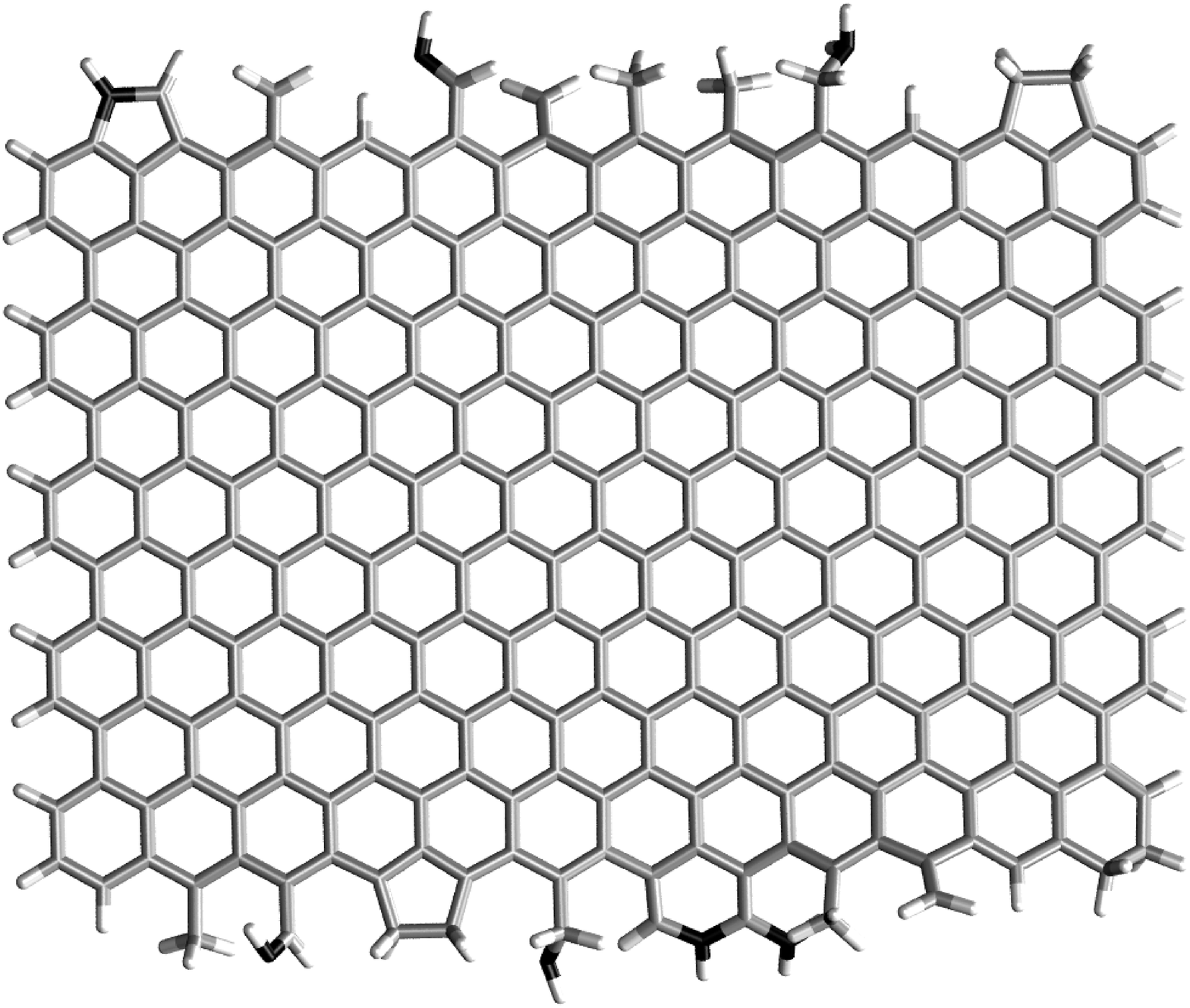}}
    \caption{(5,5) CNT functionalized with 12 pyrrolidines.  (a) initial structure, then after molecular dynamics simulations at 2000 K for (b) 50 ps and (c) 70 ps. Hydrogen is white, carbon is gray and nitrogen is black.  Note that the majority of opened edge sites in (c) are Klein-type.}
  \label{unzipp}
\end{figure}

Such edges could also form when tailoring graphene sheets\cite{Tapaszto2008,Fasoli2009} or unzipping carbon nanotubes \cite{Kosynkin2009,Jiao2009}. An armchair nanotube can be opened along its axis giving a nanoribbon with either zigzag or reconstructed Klein edges depending on the cutting line (see Fig.\ref{cut_CNT}). Under vacuum conditions the zigzag terminated ribbon is 0.32-0.82 eV/\AA \; more stable than the reconstructed Klein. However with hydrogen present the reconstructed Klein terminated ribbon is 0.03-0.36 eV/\AA \; more stable than the zigzag, the precise energy difference depending on the edge configurations \cite{supmat}.\\
To better explore the edge structures that can form during tube unzipping under experimentally relevant conditions, we next simulated pyrrolidine functionalized CNTs, since they have been found to unzip in vacuum \cite{Paiva2010}. A pyrrolidine group is formed when an azomethine ylide (CH$_{2}$NHCH$_{2}$) bonds to two C atoms of the tube \cite{Melle-Franco2004}. We modelled a 3 nm long (5,5) carbon nanotube functionalized with 12 pyrrolidine groups perpendicular to the tube axis (Fig.\ref{unzipp} (a)). The number and position of the functional groups was chosen to promote the unzipping. Fig.\ref{unzipp}(a) shows the starting structure, which is then heated to 2000 K. After 50 ps the nanotube is fully unzipped Fig.\ref{unzipp}(b). A few atoms no longer chemically bonded to the tube are taken out of the simulation. The system is then relaxed and run for 20 ps more to produce the final flat nanoribbon structure, Fig.\ref{unzipp} (c). Of the 24 \textit{unzipped} C atoms at the new boundaries, 19 show Klein edges (4 reconstructed) of which 10 C atoms have sp$^{3}$ character. A variety of Klein-related edge units can be observed along the unzipping line, including $rk_{22}$, $k_2$ and $k_3$ species. Thus these simulations suggest that Klein-based edge formation during pyrrolidine unzipping of carbon nanotubes occurs.\\

\begin{figure}
\centering
 \subfigure[]{\includegraphics[width=0.455\linewidth]{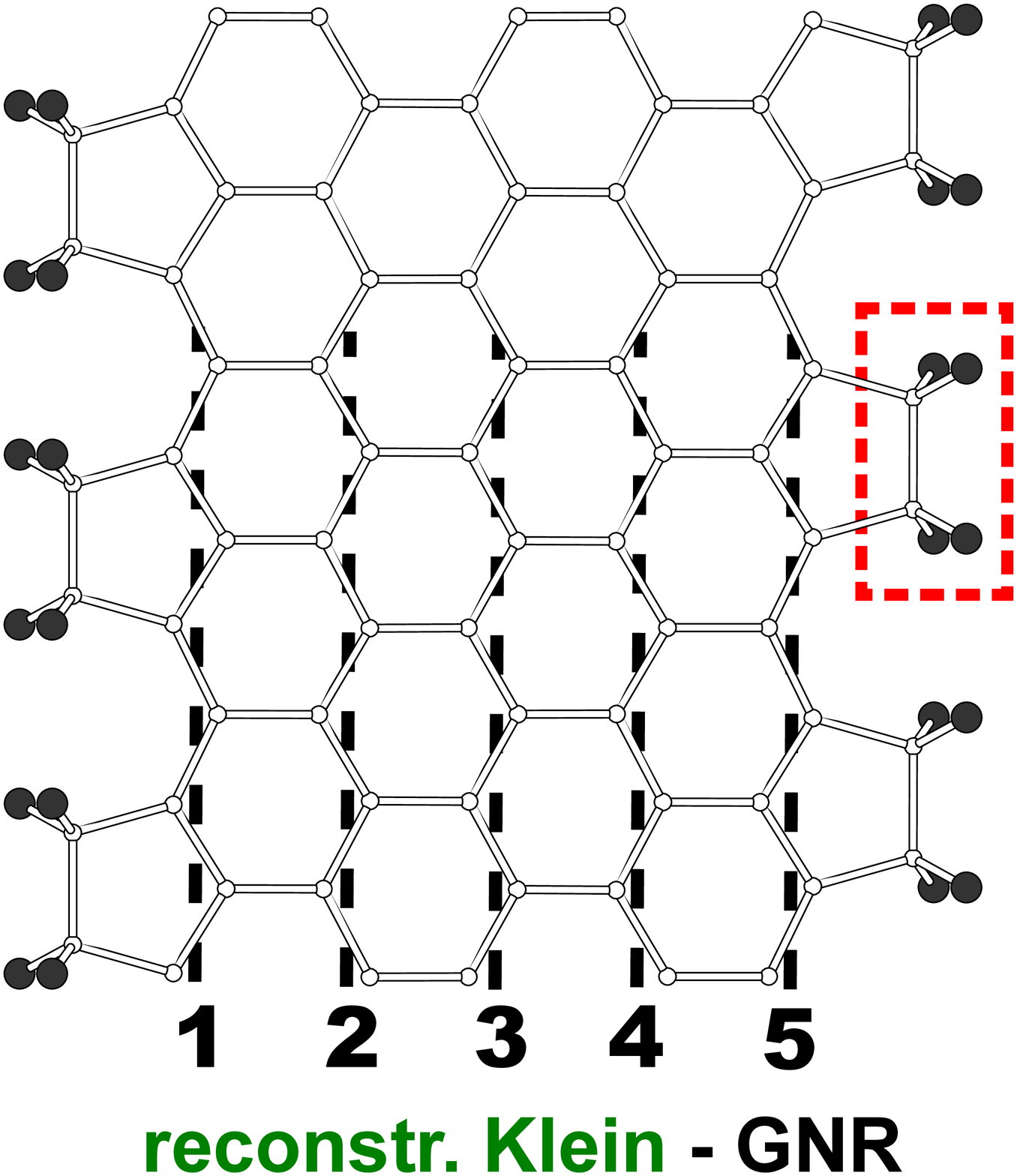}} \hspace*{0.5cm}
  \subfigure[]{\includegraphics[width=0.41\linewidth]{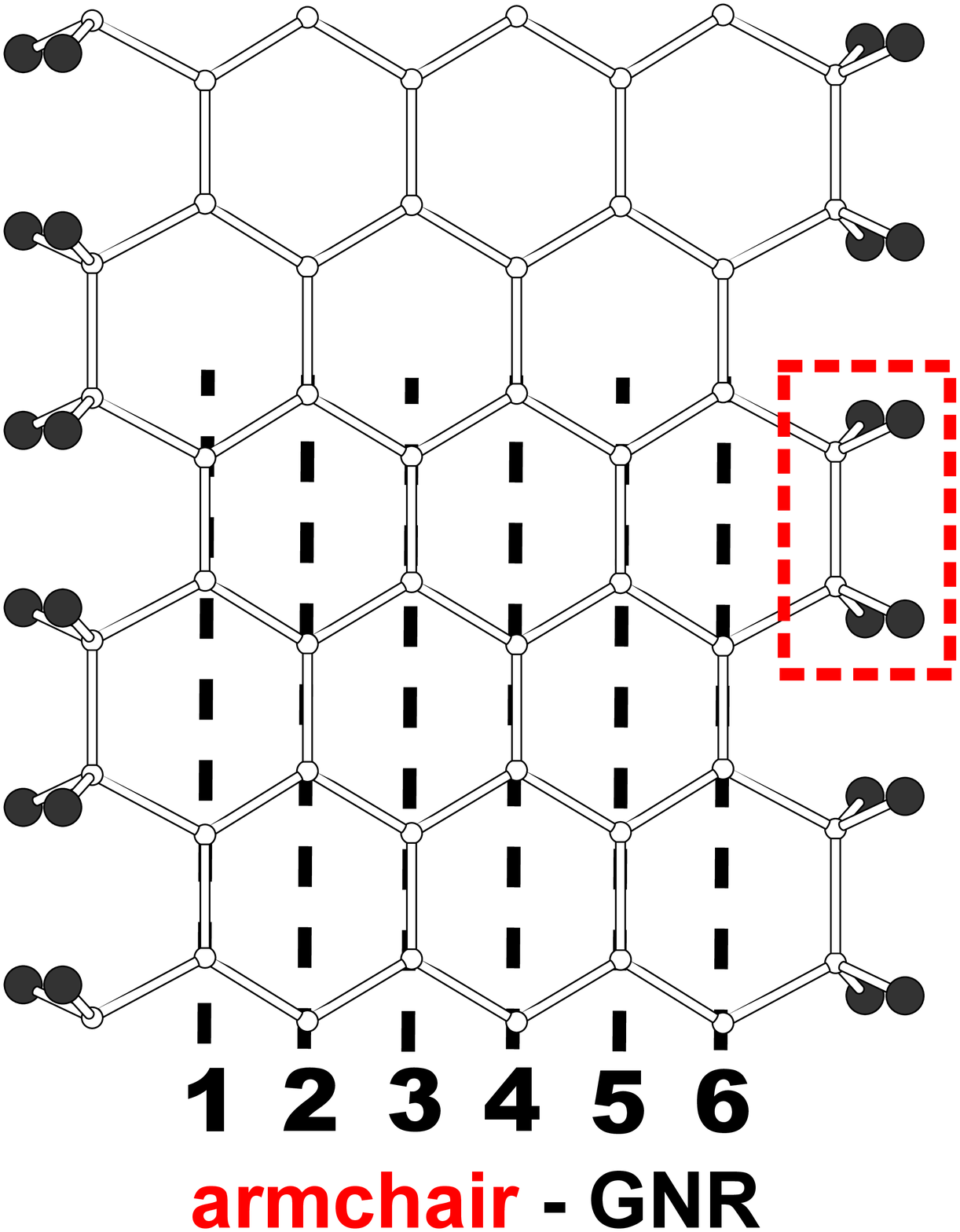}}
  \caption{(a) reconstructed Klein GNR of width $5$ with $rk_{22}$ terminated edges, (b) an $a_{22}$ terminated armchair GNR of width $6$. In both cases an ethylene C$_2$H$_4$ edge group is highlighted with a red box. On the bottom width definitions based on the sp$^2$-carbon network are marked. H atoms represented by black spheres, C atoms with white circles.}
  \label{unification}
\end{figure}

The new hydrogenated reconstructed Klein edges presented here allow us to create a unified picture of hydrogenated graphene edges. Formation energies for hydrogenated ${<}1\bar{1}00{>}$ armchair and  ${<}2\bar{1}\bar{1}0{>}$ reconstructed Klein edges are similar, with both containing C$_2$H$_4$ motifs (see Fig.\ref{unification}). It is easy to imagine intermediate chiral edges constructed similarly. For example chiral periodic edges observed experimentally from unzipped carbon nanotubes \cite{Zhang2013} appear to bear some similarity to such structures, often using intermediate temperature ($\sim$300$^{\circ}$C) H$_2$ plasma treatment to clean the graphene edges \cite{Xie2010}.\\
Reconstructed Klein edges may be difficult to observe experimentally. High resolution electron microscopy studies will rapidly detach edge hydrogen, rendering the reconstructed Klein edge unstable. Scanning tunnelling microscopy experiments under relatively low voltage bias are likely to observe a zigzag type edge. This is because the C$_2$H$_4$ reconstructed Klein edge groups are sp$^3$ coordinated with deeper electronic states, and hence will appear dark at voltage biases sufficient to image the $\pi$-bonded sp$^2$ carbon basal plane network.\\ 
Although the DFT calculations presented here concern only hydrogenation of graphene edges we anticipate a qualitatively similar stability of ${<}2\bar{1}\bar{1}0{>}$ reconstructed Klein edges with different edge functional groups\cite{Wagner2011a,Cocchi2011a}.\\

In summary, we show here that the most stable hydrogenated graphene edges along ${<}2\bar{1}\bar{1}0{>}$ are not the previously reported zigzag edge structures, but instead reconstructed Klein edge structures ($rk_{22}$ and $rk_{22}+z_2$). The stability of these edges and their derivatives approaches that of hydrogenated armchair edges. Edge methylation stabilises edges still further, but only under the condition that the edge groups can undergo significant out-of-plane edge rippling. The new stable hydrogenated edge types are predicted to occur under experimentally attainable conditions, as demonstrated via molecular dynamics simulations of pyrrolidine promoted unzipping of carbon nanotubes. The most stable edge structures are all non-magnetic. It will be important to revisit previous experimental studies, notably on graphene nanoribbon production, in light of these new edge types.\\

\begin{acknowledgments}
P. W., V. V. I. and C. P. E. thank the NANOSIM-GRAPHENE project ANR-09-NANO-016-01 funded by the French National Research Agency (ANR). P. W., B. H., J.-J. A. and C. P. E. thank the SPRINT ANR-10-BLAN-0819 project. M. M. F. thanks the Portuguese ``Funda\c{c}\~ao para a Ci\^encia e a Tecnologia'' through the program Ci\^encia 2008 and contracts PEst-OE/EEI/UI0752/2011 and CONC-REEQ/443/2005. P. R. B. thanks the CNRS for financial support. We thank COST project MP0901 NanoTP for support.

\end{acknowledgments}


\begin{thebibliography}{47}%
\makeatletter
\providecommand \@ifxundefined [1]{%
 \@ifx{#1\undefined}
}%
\providecommand \@ifnum [1]{%
 \ifnum #1\expandafter \@firstoftwo
 \else \expandafter \@secondoftwo
 \fi
}%
\providecommand \@ifx [1]{%
 \ifx #1\expandafter \@firstoftwo
 \else \expandafter \@secondoftwo
 \fi
}%
\providecommand \natexlab [1]{#1}%
\providecommand \enquote  [1]{``#1''}%
\providecommand \bibnamefont  [1]{#1}%
\providecommand \bibfnamefont [1]{#1}%
\providecommand \citenamefont [1]{#1}%
\providecommand \href@noop [0]{\@secondoftwo}%
\providecommand \href [0]{\begingroup \@sanitize@url \@href}%
\providecommand \@href[1]{\@@startlink{#1}\@@href}%
\providecommand \@@href[1]{\endgroup#1\@@endlink}%
\providecommand \@sanitize@url [0]{\catcode `\\12\catcode `\$12\catcode
  `\&12\catcode `\#12\catcode `\^12\catcode `\_12\catcode `\%12\relax}%
\providecommand \@@startlink[1]{}%
\providecommand \@@endlink[0]{}%
\providecommand \url  [0]{\begingroup\@sanitize@url \@url }%
\providecommand \@url [1]{\endgroup\@href {#1}{\urlprefix }}%
\providecommand \urlprefix  [0]{URL }%
\providecommand \Eprint [0]{\href }%
\@ifxundefined \urlstyle {%
  \providecommand \doi  [0]{\begingroup \@sanitize@url \@doi}%
  \providecommand \@doi [1]{\endgroup \@@startlink {\doibase
  #1}doi:\discretionary {}{}{}#1\@@endlink }%
}{%
  \providecommand \doi  [0]{doi:\discretionary{}{}{}\begingroup
  \urlstyle{rm}\Url }%
}%
\providecommand \doibase [0]{http://dx.doi.org/}%
\providecommand \Doi [0]{\begingroup \@sanitize@url \@Doi }%
\providecommand \@Doi  [1]{\endgroup\@@startlink{\doibase#1}\@@Doi}%
\providecommand \@@Doi [1]{#1\@@endlink}%
\providecommand \selectlanguage [0]{\@gobble}%
\providecommand \bibinfo  [0]{\@secondoftwo}%
\providecommand \bibfield  [0]{\@secondoftwo}%
\providecommand \translation [1]{[#1]}%
\providecommand \BibitemOpen [0]{}%
\providecommand \bibitemStop [0]{}%
\providecommand \bibitemNoStop [0]{.\EOS\space}%
\providecommand \EOS [0]{\spacefactor3000\relax}%
\providecommand \BibitemShut  [1]{\csname bibitem#1\endcsname}%
%</preamble>
\bibitem [{\citenamefont {Nakada}\ \emph {et~al.}(1996)\citenamefont {Nakada},
  \citenamefont {Fujita}, \citenamefont {Dresselhaus},\ and\ \citenamefont
  {Dresselhaus}}]{Nakada1996}%
  \BibitemOpen
  \bibfield  {author} {\bibinfo {author} {\bibfnamefont {K.}~\bibnamefont
  {Nakada}}, \bibinfo {author} {\bibfnamefont {M.}~\bibnamefont {Fujita}},
  \bibinfo {author} {\bibfnamefont {G.}~\bibnamefont {Dresselhaus}}, \ and\
  \bibinfo {author} {\bibfnamefont {M.~S.}\ \bibnamefont {Dresselhaus}},\
  }\href {http://dx.doi.org/10.1103/PhysRevB.54.17954} {\bibfield  {journal}
  {\bibinfo  {journal} {Phys. Rev. B},\ }\textbf {\bibinfo {volume} {54}},\
  \bibinfo {pages} {17954} (\bibinfo {year} {1996})}\BibitemShut {NoStop}%
\bibitem [{\citenamefont {Wakabayashi}\ \emph {et~al.}(1999)\citenamefont
  {Wakabayashi}, \citenamefont {Fujita}, \citenamefont {Ajiki},\ and\
  \citenamefont {Sigrist}}]{Wakabayashi1999}%
  \BibitemOpen
  \bibfield  {author} {\bibinfo {author} {\bibfnamefont {K.}~\bibnamefont
  {Wakabayashi}}, \bibinfo {author} {\bibfnamefont {M.}~\bibnamefont {Fujita}},
  \bibinfo {author} {\bibfnamefont {H.}~\bibnamefont {Ajiki}}, \ and\ \bibinfo
  {author} {\bibfnamefont {M.}~\bibnamefont {Sigrist}},\ }\href
  {http://dx.doi.org/10.1103/PhysRevB.59.8271} {\bibfield  {journal} {\bibinfo
  {journal} {Phys. Rev. B},\ }\textbf {\bibinfo {volume} {59}},\ \bibinfo
  {pages} {8271} (\bibinfo {year} {1999})}\BibitemShut {NoStop}%
\bibitem [{\citenamefont {Can\c{c}ado}\ \emph {et~al.}(2004)\citenamefont
  {Can\c{c}ado}, \citenamefont {Pimenta}, \citenamefont {Neves}, \citenamefont
  {Dantas},\ and\ \citenamefont {Jorio}}]{Cancado2004}%
  \BibitemOpen
  \bibfield  {author} {\bibinfo {author} {\bibfnamefont {L.~G.}\ \bibnamefont
  {Can\c{c}ado}}, \bibinfo {author} {\bibfnamefont {M.~A.}\ \bibnamefont
  {Pimenta}}, \bibinfo {author} {\bibfnamefont {B.~R.~A.}\ \bibnamefont
  {Neves}}, \bibinfo {author} {\bibfnamefont {M.~S.~S.}\ \bibnamefont
  {Dantas}}, \ and\ \bibinfo {author} {\bibfnamefont {A.}~\bibnamefont
  {Jorio}},\ }\href {http://dx.doi.org/10.1103/PhysRevLett.93.247401}
  {\bibfield  {journal} {\bibinfo  {journal} {Phys. Rev. Lett.},\ }\textbf
  {\bibinfo {volume} {93}},\ \bibinfo {pages} {247401} (\bibinfo {year}
  {2004})}\BibitemShut {NoStop}%
\bibitem [{\citenamefont {Geim}\ and\ \citenamefont
  {Novoselov}(2007)}]{Geim2007}%
  \BibitemOpen
  \bibfield  {author} {\bibinfo {author} {\bibfnamefont {A.~K.}\ \bibnamefont
  {Geim}}\ and\ \bibinfo {author} {\bibfnamefont {K.~S.}\ \bibnamefont
  {Novoselov}},\ }\href {http://dx.doi.org/10.1038/nmat1849} {\bibfield
  {journal} {\bibinfo  {journal} {Nature Mater.},\ }\textbf {\bibinfo {volume}
  {6}},\ \bibinfo {pages} {183} (\bibinfo {year} {2007})}\BibitemShut {NoStop}%
\bibitem [{\citenamefont {Cervantes-Sodi}\ \emph {et~al.}(2008)\citenamefont
  {Cervantes-Sodi}, \citenamefont {Cs\'anyi}, \citenamefont {Piscanec},\ and\
  \citenamefont {Ferrari}}]{CervantesSodi2008}%
  \BibitemOpen
  \bibfield  {author} {\bibinfo {author} {\bibfnamefont {F.}~\bibnamefont
  {Cervantes-Sodi}}, \bibinfo {author} {\bibfnamefont {G.}~\bibnamefont
  {Cs\'anyi}}, \bibinfo {author} {\bibfnamefont {S.}~\bibnamefont {Piscanec}},
  \ and\ \bibinfo {author} {\bibfnamefont {A.~C.}\ \bibnamefont {Ferrari}},\
  }\href {http://dx.doi.org/10.1103/PhysRevB.77.165427} {\bibfield  {journal}
  {\bibinfo  {journal} {Phys. Rev. B},\ }\textbf {\bibinfo {volume} {77}},\
  \bibinfo {pages} {165427} (\bibinfo {year} {2008})}\BibitemShut {NoStop}%
\bibitem [{\citenamefont {Wassmann}\ \emph {et~al.}(2008)\citenamefont
  {Wassmann}, \citenamefont {Seitsonen}, \citenamefont {Saitta}, \citenamefont
  {Lazzeri},\ and\ \citenamefont {Mauri}}]{Wassmann2008}%
  \BibitemOpen
  \bibfield  {author} {\bibinfo {author} {\bibfnamefont {T.}~\bibnamefont
  {Wassmann}}, \bibinfo {author} {\bibfnamefont {A.~P.}\ \bibnamefont
  {Seitsonen}}, \bibinfo {author} {\bibfnamefont {A.~M.}\ \bibnamefont
  {Saitta}}, \bibinfo {author} {\bibfnamefont {M.}~\bibnamefont {Lazzeri}}, \
  and\ \bibinfo {author} {\bibfnamefont {F.}~\bibnamefont {Mauri}},\ }\href
  {http://dx.doi.org/10.1103/PhysRevLett.101.096402} {\bibfield  {journal}
  {\bibinfo  {journal} {Phys. Rev. Lett.},\ }\textbf {\bibinfo {volume}
  {101}},\ \bibinfo {pages} {096402} (\bibinfo {year} {2008})}\BibitemShut
  {NoStop}%
\bibitem [{\citenamefont {Koskinen}\ \emph {et~al.}(2008)\citenamefont
  {Koskinen}, \citenamefont {Malola},\ and\ \citenamefont
  {H\"akkinen}}]{Koskinen2008}%
  \BibitemOpen
  \bibfield  {author} {\bibinfo {author} {\bibfnamefont {P.}~\bibnamefont
  {Koskinen}}, \bibinfo {author} {\bibfnamefont {S.}~\bibnamefont {Malola}}, \
  and\ \bibinfo {author} {\bibfnamefont {H.}~\bibnamefont {H\"akkinen}},\
  }\href {http://dx.doi.org/10.1103/PhysRevLett.101.115502} {\bibfield
  {journal} {\bibinfo  {journal} {Phys. Rev. Lett.},\ }\textbf {\bibinfo
  {volume} {101}},\ \bibinfo {pages} {115502} (\bibinfo {year}
  {2008})}\BibitemShut {NoStop}%
\bibitem [{\citenamefont {Girit}\ \emph {et~al.}(2009)\citenamefont {Girit},
  \citenamefont {Meyer}, \citenamefont {Erni}, \citenamefont {Rossell},
  \citenamefont {Kisielowski}, \citenamefont {Yang}, \citenamefont {Park},
  \citenamefont {Crommie}, \citenamefont {Cohen}, \citenamefont {Louie},\ and\
  \citenamefont {Zettl}}]{Girit2009}%
  \BibitemOpen
  \bibfield  {author} {\bibinfo {author} {\bibfnamefont {C.}~\bibnamefont
  {Girit}}, \bibinfo {author} {\bibfnamefont {J.}~\bibnamefont {Meyer}},
  \bibinfo {author} {\bibfnamefont {R.}~\bibnamefont {Erni}}, \bibinfo {author}
  {\bibfnamefont {M.}~\bibnamefont {Rossell}}, \bibinfo {author} {\bibfnamefont
  {C.}~\bibnamefont {Kisielowski}}, \bibinfo {author} {\bibfnamefont
  {L.}~\bibnamefont {Yang}}, \bibinfo {author} {\bibfnamefont {C.}~\bibnamefont
  {Park}}, \bibinfo {author} {\bibfnamefont {M.}~\bibnamefont {Crommie}},
  \bibinfo {author} {\bibfnamefont {M.}~\bibnamefont {Cohen}}, \bibinfo
  {author} {\bibfnamefont {S.}~\bibnamefont {Louie}}, \ and\ \bibinfo {author}
  {\bibfnamefont {A.}~\bibnamefont {Zettl}},\ }\href
  {http://dx.doi.org/10.1126/science.1166999} {\bibfield  {journal} {\bibinfo
  {journal} {Science},\ }\textbf {\bibinfo {volume} {323}},\ \bibinfo {pages}
  {1705} (\bibinfo {year} {2009})}\BibitemShut {NoStop}%
\bibitem [{\citenamefont {Jia}\ \emph {et~al.}(2011)\citenamefont {Jia},
  \citenamefont {Campos-Delgado}, \citenamefont {Terrones}, \citenamefont
  {Meunier},\ and\ \citenamefont {Dresselhaus}}]{Jia2011}%
  \BibitemOpen
  \bibfield  {author} {\bibinfo {author} {\bibfnamefont {X.}~\bibnamefont
  {Jia}}, \bibinfo {author} {\bibfnamefont {J.}~\bibnamefont {Campos-Delgado}},
  \bibinfo {author} {\bibfnamefont {M.}~\bibnamefont {Terrones}}, \bibinfo
  {author} {\bibfnamefont {V.}~\bibnamefont {Meunier}}, \ and\ \bibinfo
  {author} {\bibfnamefont {M.~S.}\ \bibnamefont {Dresselhaus}},\ }\href
  {http://dx.doi.org/10.1039/c0nr00600a} {\bibfield  {journal} {\bibinfo
  {journal} {Nanoscale},\ }\textbf {\bibinfo {volume} {3}},\ \bibinfo {pages}
  {86} (\bibinfo {year} {2011})}\BibitemShut {NoStop}%
\bibitem [{\citenamefont {Ritter}\ and\ \citenamefont
  {Lyding}(2009)}]{Ritter2009}%
  \BibitemOpen
  \bibfield  {author} {\bibinfo {author} {\bibfnamefont {K.~A.}\ \bibnamefont
  {Ritter}}\ and\ \bibinfo {author} {\bibfnamefont {J.~W.}\ \bibnamefont
  {Lyding}},\ }\href {http://dx.doi.org/10.1038/nmat2378} {\bibfield  {journal}
  {\bibinfo  {journal} {Nature Mater.},\ }\textbf {\bibinfo {volume} {8}},\
  \bibinfo {pages} {235} (\bibinfo {year} {2009})}\BibitemShut {NoStop}%
\bibitem [{\citenamefont {Gan}\ and\ \citenamefont
  {Srolovitz}(2010)}]{Gan2010}%
  \BibitemOpen
  \bibfield  {author} {\bibinfo {author} {\bibfnamefont {C.~K.}\ \bibnamefont
  {Gan}}\ and\ \bibinfo {author} {\bibfnamefont {D.~J.}\ \bibnamefont
  {Srolovitz}},\ }\href {http://dx.doi.org/10.1103/PhysRevB.81.125445}
  {\bibfield  {journal} {\bibinfo  {journal} {Phys. Rev. B},\ }\textbf
  {\bibinfo {volume} {81}},\ \bibinfo {pages} {125445} (\bibinfo {year}
  {2010})}\BibitemShut {NoStop}%
\bibitem [{\citenamefont {Wagner}\ \emph {et~al.}(2011)\citenamefont {Wagner},
  \citenamefont {Ewels}, \citenamefont {Ivanovskaya}, \citenamefont {Briddon},
  \citenamefont {Pateau},\ and\ \citenamefont {Humbert}}]{Wagner2011a}%
  \BibitemOpen
  \bibfield  {author} {\bibinfo {author} {\bibfnamefont {P.}~\bibnamefont
  {Wagner}}, \bibinfo {author} {\bibfnamefont {C.~P.}\ \bibnamefont {Ewels}},
  \bibinfo {author} {\bibfnamefont {V.~V.}\ \bibnamefont {Ivanovskaya}},
  \bibinfo {author} {\bibfnamefont {P.~R.}\ \bibnamefont {Briddon}}, \bibinfo
  {author} {\bibfnamefont {A.}~\bibnamefont {Pateau}}, \ and\ \bibinfo {author}
  {\bibfnamefont {B.}~\bibnamefont {Humbert}},\ }\href
  {http://dx.doi.org/10.1103/PhysRevB.84.134110} {\bibfield  {journal}
  {\bibinfo  {journal} {Phys. Rev. B},\ }\textbf {\bibinfo {volume} {84}},\
  \bibinfo {pages} {134110} (\bibinfo {year} {2011})}\BibitemShut {NoStop}%
\bibitem [{\citenamefont {Cocchi}\ \emph {et~al.}(2011)\citenamefont {Cocchi},
  \citenamefont {Prezzi}, \citenamefont {Ruini}, \citenamefont {Caldas},\ and\
  \citenamefont {Molinari}}]{Cocchi2011a}%
  \BibitemOpen
  \bibfield  {author} {\bibinfo {author} {\bibfnamefont {C.}~\bibnamefont
  {Cocchi}}, \bibinfo {author} {\bibfnamefont {D.}~\bibnamefont {Prezzi}},
  \bibinfo {author} {\bibfnamefont {A.}~\bibnamefont {Ruini}}, \bibinfo
  {author} {\bibfnamefont {M.~J.}\ \bibnamefont {Caldas}}, \ and\ \bibinfo
  {author} {\bibfnamefont {E.}~\bibnamefont {Molinari}},\ }\href
  {http://dx.doi.org/10.1021/jz200472a} {\bibfield  {journal} {\bibinfo
  {journal} {J. Phys. Chem. Lett.},\ }\textbf {\bibinfo {volume} {2}},\
  \bibinfo {pages} {1315} (\bibinfo {year} {2011})}\BibitemShut {NoStop}%
\bibitem [{\citenamefont {Branicio}\ \emph {et~al.}(2011)\citenamefont
  {Branicio}, \citenamefont {Jhon}, \citenamefont {Gan},\ and\ \citenamefont
  {Srolovitz}}]{Branicio2011}%
  \BibitemOpen
  \bibfield  {author} {\bibinfo {author} {\bibfnamefont {P.~S.}\ \bibnamefont
  {Branicio}}, \bibinfo {author} {\bibfnamefont {M.~H.}\ \bibnamefont {Jhon}},
  \bibinfo {author} {\bibfnamefont {C.~K.}\ \bibnamefont {Gan}}, \ and\
  \bibinfo {author} {\bibfnamefont {D.~J.}\ \bibnamefont {Srolovitz}},\ }\href
  {http://dx.doi.org/10.1088/0965-0393/19/5/054002} {\bibfield  {journal}
  {\bibinfo  {journal} {Modell. Simul. Mater. Science Engineering},\ }\textbf
  {\bibinfo {volume} {19}},\ \bibinfo {pages} {054002} (\bibinfo {year}
  {2011})}\BibitemShut {NoStop}%
\bibitem [{\citenamefont {Ivanovskaya}\ \emph {et~al.}(2011)\citenamefont
  {Ivanovskaya}, \citenamefont {Zobelli}, \citenamefont {Wagner}, \citenamefont
  {Heggie}, \citenamefont {Briddon}, \citenamefont {Rayson},\ and\
  \citenamefont {Ewels}}]{Ivanovskaya2011}%
  \BibitemOpen
  \bibfield  {author} {\bibinfo {author} {\bibfnamefont {V.~V.}\ \bibnamefont
  {Ivanovskaya}}, \bibinfo {author} {\bibfnamefont {A.}~\bibnamefont
  {Zobelli}}, \bibinfo {author} {\bibfnamefont {P.}~\bibnamefont {Wagner}},
  \bibinfo {author} {\bibfnamefont {M.~I.}\ \bibnamefont {Heggie}}, \bibinfo
  {author} {\bibfnamefont {P.~R.}\ \bibnamefont {Briddon}}, \bibinfo {author}
  {\bibfnamefont {M.~J.}\ \bibnamefont {Rayson}}, \ and\ \bibinfo {author}
  {\bibfnamefont {C.~P.}\ \bibnamefont {Ewels}},\ }\href
  {http://dx.doi.org/10.1103/PhysRevLett.107.065502} {\bibfield  {journal}
  {\bibinfo  {journal} {Phys. Rev. Lett.},\ }\textbf {\bibinfo {volume}
  {107}},\ \bibinfo {pages} {065502} (\bibinfo {year} {2011})}\BibitemShut
  {NoStop}%
\bibitem [{\citenamefont {Klein}(1994)}]{Klein1994}%
  \BibitemOpen
  \bibfield  {author} {\bibinfo {author} {\bibfnamefont {D.}~\bibnamefont
  {Klein}},\ }\href {http://dx.doi.org/10.1016/0009-2614(93)E1378-T} {\bibfield
   {journal} {\bibinfo  {journal} {Chem. Phys. Lett.},\ }\textbf {\bibinfo
  {volume} {217}},\ \bibinfo {pages} {261} (\bibinfo {year}
  {1994})}\BibitemShut {NoStop}%
\bibitem [{\citenamefont {Zobelli}\ \emph {et~al.}(2012)\citenamefont
  {Zobelli}, \citenamefont {Ivanovskaya}, \citenamefont {Wagner}, \citenamefont
  {{Suarez-Martinez}}, \citenamefont {Yaya},\ and\ \citenamefont
  {Ewels}}]{Zobelli2012}%
  \BibitemOpen
  \bibfield  {author} {\bibinfo {author} {\bibfnamefont {A.}~\bibnamefont
  {Zobelli}}, \bibinfo {author} {\bibfnamefont {V.}~\bibnamefont
  {Ivanovskaya}}, \bibinfo {author} {\bibfnamefont {P.}~\bibnamefont {Wagner}},
  \bibinfo {author} {\bibfnamefont {I.}~\bibnamefont {{Suarez-Martinez}}},
  \bibinfo {author} {\bibfnamefont {A.}~\bibnamefont {Yaya}}, \ and\ \bibinfo
  {author} {\bibfnamefont {C.~P.}\ \bibnamefont {Ewels}},\ }\href
  {http://dx.doi.org/10.1002/pssb.201100630} {\bibfield  {journal} {\bibinfo
  {journal} {Phys. Status Solidi B},\ }\textbf {\bibinfo {volume} {249}},\
  \bibinfo {pages} {276} (\bibinfo {year} {2012})}\BibitemShut {NoStop}%
\bibitem [{\citenamefont {Kunstmann}\ \emph {et~al.}(2011)\citenamefont
  {Kunstmann}, \citenamefont {Ozdo{\u{g}}an}, \citenamefont {Quandt},\ and\
  \citenamefont {Fehske}}]{Kunstmann2011}%
  \BibitemOpen
  \bibfield  {author} {\bibinfo {author} {\bibfnamefont {J.}~\bibnamefont
  {Kunstmann}}, \bibinfo {author} {\bibfnamefont {C.}~\bibnamefont
  {Ozdo{\u{g}}an}}, \bibinfo {author} {\bibfnamefont {A.}~\bibnamefont
  {Quandt}}, \ and\ \bibinfo {author} {\bibfnamefont {H.}~\bibnamefont
  {Fehske}},\ }\href {http://dx.doi.org/10.1103/PhysRevB.83.045414} {\bibfield
  {journal} {\bibinfo  {journal} {Phys. Rev. B},\ }\textbf {\bibinfo {volume}
  {83}},\ \bibinfo {pages} {045414} (\bibinfo {year} {2011})}\BibitemShut
  {NoStop}%
\bibitem [{\citenamefont {Briddon}\ and\ \citenamefont
  {Jones}(2000)}]{Briddon2000}%
  \BibitemOpen
  \bibfield  {author} {\bibinfo {author} {\bibfnamefont {P.~R.}\ \bibnamefont
  {Briddon}}\ and\ \bibinfo {author} {\bibfnamefont {R.}~\bibnamefont
  {Jones}},\ }\href
  {http://dx.doi.org/10.1002/(SICI)1521-3951(200001)217:1<131::AID-PSSB131>3.0%
.CO;2-M} {\bibfield  {journal} {\bibinfo  {journal} {Phys. Status Solidi B},\
  }\textbf {\bibinfo {volume} {217}},\ \bibinfo {pages} {131} (\bibinfo {year}
  {2000})}\BibitemShut {NoStop}%
\bibitem [{\citenamefont {Rayson}\ and\ \citenamefont
  {Briddon}(2009)}]{Rayson2009}%
  \BibitemOpen
  \bibfield  {author} {\bibinfo {author} {\bibfnamefont {M.~J.}\ \bibnamefont
  {Rayson}}\ and\ \bibinfo {author} {\bibfnamefont {P.~R.}\ \bibnamefont
  {Briddon}},\ }\href {http://dx.doi.org/10.1103/PhysRevB.80.205104} {\bibfield
   {journal} {\bibinfo  {journal} {Phys. Rev. B},\ }\textbf {\bibinfo {volume}
  {80}},\ \bibinfo {pages} {205104} (\bibinfo {year} {2009})}\BibitemShut
  {NoStop}%
\bibitem [{\citenamefont {Briddon}\ and\ \citenamefont
  {Rayson}(2011)}]{Briddon2011}%
  \BibitemOpen
  \bibfield  {author} {\bibinfo {author} {\bibfnamefont {P.~R.}\ \bibnamefont
  {Briddon}}\ and\ \bibinfo {author} {\bibfnamefont {M.~J.}\ \bibnamefont
  {Rayson}},\ }\href {http://dx.doi.org/10.1002/pssb.201046147} {\bibfield
  {journal} {\bibinfo  {journal} {Phys. Status Solidi B},\ }\textbf {\bibinfo
  {volume} {248}},\ \bibinfo {pages} {1309} (\bibinfo {year}
  {2011})}\BibitemShut {NoStop}%
\bibitem [{\citenamefont {Hartwigsen}\ \emph {et~al.}(1998)\citenamefont
  {Hartwigsen}, \citenamefont {Goedecker},\ and\ \citenamefont
  {Hutter}}]{Hartwigsen1998}%
  \BibitemOpen
  \bibfield  {author} {\bibinfo {author} {\bibfnamefont {C.}~\bibnamefont
  {Hartwigsen}}, \bibinfo {author} {\bibfnamefont {S.}~\bibnamefont
  {Goedecker}}, \ and\ \bibinfo {author} {\bibfnamefont {J.}~\bibnamefont
  {Hutter}},\ }\href {http://dx.doi.org/10.1103/PhysRevB.58.3641} {\bibfield
  {journal} {\bibinfo  {journal} {Phys. Rev. B},\ }\textbf {\bibinfo {volume}
  {58}},\ \bibinfo {pages} {3641} (\bibinfo {year} {1998})}\BibitemShut
  {NoStop}%
\bibitem [{\citenamefont {Elstner}\ \emph {et~al.}(1998)\citenamefont
  {Elstner}, \citenamefont {Porezag}, \citenamefont {Jungnickel}, \citenamefont
  {Elsner}, \citenamefont {Haugk}, \citenamefont {Frauenheim}, \citenamefont
  {Suhai},\ and\ \citenamefont {Seifert}}]{Elstner1998}%
  \BibitemOpen
  \bibfield  {author} {\bibinfo {author} {\bibfnamefont {M.}~\bibnamefont
  {Elstner}}, \bibinfo {author} {\bibfnamefont {D.}~\bibnamefont {Porezag}},
  \bibinfo {author} {\bibfnamefont {G.}~\bibnamefont {Jungnickel}}, \bibinfo
  {author} {\bibfnamefont {J.}~\bibnamefont {Elsner}}, \bibinfo {author}
  {\bibfnamefont {M.}~\bibnamefont {Haugk}}, \bibinfo {author} {\bibfnamefont
  {T.}~\bibnamefont {Frauenheim}}, \bibinfo {author} {\bibfnamefont
  {S.}~\bibnamefont {Suhai}}, \ and\ \bibinfo {author} {\bibfnamefont
  {G.}~\bibnamefont {Seifert}},\ }\href
  {http://ldx.doi.org/10.1103/PhysRevB.58.7260} {\bibfield  {journal} {\bibinfo
   {journal} {Phys. Rev. B},\ }\textbf {\bibinfo {volume} {58}},\ \bibinfo
  {pages} {7260} (\bibinfo {year} {1998})}\BibitemShut {NoStop}%
\bibitem [{\citenamefont {Aradi}\ \emph {et~al.}(2007)\citenamefont {Aradi},
  \citenamefont {Hourahine},\ and\ \citenamefont {Frauenheim}}]{Aradi2007}%
  \BibitemOpen
  \bibfield  {author} {\bibinfo {author} {\bibfnamefont {B.}~\bibnamefont
  {Aradi}}, \bibinfo {author} {\bibfnamefont {B.}~\bibnamefont {Hourahine}}, \
  and\ \bibinfo {author} {\bibfnamefont {T.}~\bibnamefont {Frauenheim}},\
  }\href {http://dx.doi.org/10.1021/jp070186p} {\bibfield  {journal} {\bibinfo
  {journal} {J. Phys. Chem. A},\ }\textbf {\bibinfo {volume} {111}},\ \bibinfo
  {pages} {5678} (\bibinfo {year} {2007})}\BibitemShut {NoStop}%
\bibitem [{\citenamefont {Lu}\ \emph {et~al.}(2009)\citenamefont {Lu},
  \citenamefont {Wu}, \citenamefont {Shen}, \citenamefont {Yang}, \citenamefont
  {Sha}, \citenamefont {Cai}, \citenamefont {He},\ and\ \citenamefont
  {Feng}}]{Lu2009}%
  \BibitemOpen
  \bibfield  {author} {\bibinfo {author} {\bibfnamefont {Y.~H.}\ \bibnamefont
  {Lu}}, \bibinfo {author} {\bibfnamefont {R.~Q.}\ \bibnamefont {Wu}}, \bibinfo
  {author} {\bibfnamefont {L.}~\bibnamefont {Shen}}, \bibinfo {author}
  {\bibfnamefont {M.}~\bibnamefont {Yang}}, \bibinfo {author} {\bibfnamefont
  {Z.~D.}\ \bibnamefont {Sha}}, \bibinfo {author} {\bibfnamefont {Y.~Q.}\
  \bibnamefont {Cai}}, \bibinfo {author} {\bibfnamefont {P.~M.}\ \bibnamefont
  {He}}, \ and\ \bibinfo {author} {\bibfnamefont {Y.~P.}\ \bibnamefont
  {Feng}},\ }\href {http://apl.aip.org/resource/1/applab/v94/i12/p122111_s1}
  {\bibfield  {journal} {\bibinfo  {journal} {Appl. Phys. Lett.},\ }\textbf
  {\bibinfo {volume} {94}},\ \bibinfo {pages} {122111} (\bibinfo {year}
  {2009})}\BibitemShut {NoStop}%
\bibitem [{sup()}]{supmat}%
  \BibitemOpen
  \href@noop {} {\bibinfo  {journal} {Supplementary Materials available at
  http://XXX for additional information.}}\BibitemShut {Stop}%
\bibitem [{\citenamefont {Lee}\ \emph {et~al.}(2005)\citenamefont {Lee},
  \citenamefont {Son}, \citenamefont {Park}, \citenamefont {Han},\ and\
  \citenamefont {Yu}}]{Lee2005}%
  \BibitemOpen
\bibfield  {journal} {  }\bibfield  {author} {\bibinfo {author} {\bibfnamefont
  {H.}~\bibnamefont {Lee}}, \bibinfo {author} {\bibfnamefont {Y.-W.}\
  \bibnamefont {Son}}, \bibinfo {author} {\bibfnamefont {N.}~\bibnamefont
  {Park}}, \bibinfo {author} {\bibfnamefont {S.}~\bibnamefont {Han}}, \ and\
  \bibinfo {author} {\bibfnamefont {J.}~\bibnamefont {Yu}},\ }\href
  {http://dx.doi.org/10.1103/PhysRevB.72.174431} {\bibfield  {journal}
  {\bibinfo  {journal} {Phys. Rev. B},\ }\textbf {\bibinfo {volume} {72}},\
  \bibinfo {pages} {174431} (\bibinfo {year} {2005})}\BibitemShut {NoStop}%
\bibitem [{\citenamefont {Son}\ \emph {et~al.}(2006)\citenamefont {Son},
  \citenamefont {Cohen},\ and\ \citenamefont {Louie}}]{Son2006a}%
  \BibitemOpen
  \bibfield  {author} {\bibinfo {author} {\bibfnamefont {Y.-W.}\ \bibnamefont
  {Son}}, \bibinfo {author} {\bibfnamefont {M.~L.}\ \bibnamefont {Cohen}}, \
  and\ \bibinfo {author} {\bibfnamefont {S.~G.}\ \bibnamefont {Louie}},\ }\href
  {http://dx.doi.org/10.1038/nature05180} {\bibfield  {journal} {\bibinfo
  {journal} {Nature},\ }\textbf {\bibinfo {volume} {444}},\ \bibinfo {pages}
  {347} (\bibinfo {year} {2006})}\BibitemShut {NoStop}%
\bibitem [{\citenamefont {Lee}\ and\ \citenamefont {Cho}(2009)}]{Lee2009a}%
  \BibitemOpen
  \bibfield  {author} {\bibinfo {author} {\bibfnamefont {G.}~\bibnamefont
  {Lee}}\ and\ \bibinfo {author} {\bibfnamefont {K.}~\bibnamefont {Cho}},\
  }\href {http://dx.doi.org/10.1103/PhysRevB.79.165440} {\bibfield  {journal}
  {\bibinfo  {journal} {Phys. Rev. B},\ }\textbf {\bibinfo {volume} {79}},\
  \bibinfo {pages} {165440} (\bibinfo {year} {2009})}\BibitemShut {NoStop}%
\bibitem [{\citenamefont {Yazyev}(2010)}]{Yazyev2010}%
  \BibitemOpen
  \bibfield  {author} {\bibinfo {author} {\bibfnamefont {O.~V.}\ \bibnamefont
  {Yazyev}},\ }\href {http://dx.doi.org/10.1088/0034-4885/73/5/056501}
  {\bibfield  {journal} {\bibinfo  {journal} {Rep. Prog. Phys.},\ }\textbf
  {\bibinfo {volume} {73}},\ \bibinfo {pages} {056501} (\bibinfo {year}
  {2010})}\BibitemShut {NoStop}%
\bibitem [{\citenamefont {Chia}\ and\ \citenamefont {Crespi}(2012)}]{Chia2012}%
  \BibitemOpen
  \bibfield  {author} {\bibinfo {author} {\bibfnamefont {C.-I.}\ \bibnamefont
  {Chia}}\ and\ \bibinfo {author} {\bibfnamefont {V.~H.}\ \bibnamefont
  {Crespi}},\ }\href {http://dx.doi.org/10.1103/PhysRevLett.109.076802}
  {\bibfield  {journal} {\bibinfo  {journal} {Phys. Rev. Lett.},\ }\textbf
  {\bibinfo {volume} {109}},\ \bibinfo {pages} {076802} (\bibinfo {year}
  {2012})}\BibitemShut {NoStop}%
\bibitem [{\citenamefont {Celebi}\ \emph {et~al.}(2013)\citenamefont {Celebi},
  \citenamefont {Cole}, \citenamefont {Choi}, \citenamefont {Wyczisk},
  \citenamefont {Legagneux}, \citenamefont {Rupesinghe}, \citenamefont
  {Robertson}, \citenamefont {Teo},\ and\ \citenamefont {Park}}]{Celebi2013}%
  \BibitemOpen
  \bibfield  {author} {\bibinfo {author} {\bibfnamefont {K.}~\bibnamefont
  {Celebi}}, \bibinfo {author} {\bibfnamefont {M.~T.}\ \bibnamefont {Cole}},
  \bibinfo {author} {\bibfnamefont {J.~W.}\ \bibnamefont {Choi}}, \bibinfo
  {author} {\bibfnamefont {F.}~\bibnamefont {Wyczisk}}, \bibinfo {author}
  {\bibfnamefont {P.}~\bibnamefont {Legagneux}}, \bibinfo {author}
  {\bibfnamefont {N.}~\bibnamefont {Rupesinghe}}, \bibinfo {author}
  {\bibfnamefont {J.}~\bibnamefont {Robertson}}, \bibinfo {author}
  {\bibfnamefont {K.~B.~K.}\ \bibnamefont {Teo}}, \ and\ \bibinfo {author}
  {\bibfnamefont {H.~G.}\ \bibnamefont {Park}},\ }\href
  {http://dx.doi.org/10.1021/nl303934v} {\bibfield  {journal} {\bibinfo
  {journal} {Nano Lett.}} (\bibinfo {year} {2013})}\BibitemShut {NoStop}%
\bibitem [{\citenamefont {Treier}\ \emph {et~al.}(2011)\citenamefont {Treier},
  \citenamefont {Pignedoli}, \citenamefont {Laino}, \citenamefont {Rieger},
  \citenamefont {M\"ullen}, \citenamefont {Passerone},\ and\ \citenamefont
  {Fasel}}]{Treier2011}%
  \BibitemOpen
  \bibfield  {author} {\bibinfo {author} {\bibfnamefont {M.}~\bibnamefont
  {Treier}}, \bibinfo {author} {\bibfnamefont {C.~A.}\ \bibnamefont
  {Pignedoli}}, \bibinfo {author} {\bibfnamefont {T.}~\bibnamefont {Laino}},
  \bibinfo {author} {\bibfnamefont {R.}~\bibnamefont {Rieger}}, \bibinfo
  {author} {\bibfnamefont {K.}~\bibnamefont {M\"ullen}}, \bibinfo {author}
  {\bibfnamefont {D.}~\bibnamefont {Passerone}}, \ and\ \bibinfo {author}
  {\bibfnamefont {R.}~\bibnamefont {Fasel}},\ }\href
  {http://dx.doi.org/10.1038/nchem.891} {\bibfield  {journal} {\bibinfo
  {journal} {Nature Chem.},\ }\textbf {\bibinfo {volume} {3}},\ \bibinfo
  {pages} {61} (\bibinfo {year} {2011})}\BibitemShut {NoStop}%
\bibitem [{\citenamefont {Zhang}\ \emph {et~al.}(2011)\citenamefont {Zhang},
  \citenamefont {Wu}, \citenamefont {Li},\ and\ \citenamefont
  {Yang}}]{Zhang2011b}%
  \BibitemOpen
  \bibfield  {author} {\bibinfo {author} {\bibfnamefont {W.}~\bibnamefont
  {Zhang}}, \bibinfo {author} {\bibfnamefont {P.}~\bibnamefont {Wu}}, \bibinfo
  {author} {\bibfnamefont {Z.}~\bibnamefont {Li}}, \ and\ \bibinfo {author}
  {\bibfnamefont {J.}~\bibnamefont {Yang}},\ }\href
  {http://dx.doi.org/10.1021/jp2006827} {\bibfield  {journal} {\bibinfo
  {journal} {J. Phys. Chem. C},\ }\textbf {\bibinfo {volume} {115}},\ \bibinfo
  {pages} {17782} (\bibinfo {year} {2011})}\BibitemShut {NoStop}%
\bibitem [{\citenamefont {Chase}(1998)}]{Chase1998}%
  \BibitemOpen
  \bibfield  {author} {\bibinfo {author} {\bibfnamefont {M.~W.}\ \bibnamefont
  {Chase}},\ }\href {http://webbook.nist.gov/cgi/inchi/InChI%3D1S/H2/h1H}
  {\bibfield  {journal} {\bibinfo  {journal} {J. Phys. Chem. Ref. Data},\
  }\textbf {\bibinfo {volume} {9}},\ \bibinfo {pages} {1} (\bibinfo {year}
  {1998})}\BibitemShut {NoStop}%
\bibitem [{\citenamefont {Li}\ \emph {et~al.}(2009)\citenamefont {Li},
  \citenamefont {Cai}, \citenamefont {An}, \citenamefont {Kim}, \citenamefont
  {Nah}, \citenamefont {Yang}, \citenamefont {Piner}, \citenamefont
  {Velamakanni}, \citenamefont {Jung}, \citenamefont {Tutuc}, \citenamefont
  {Banerjee}, \citenamefont {Colombo},\ and\ \citenamefont {Ruoff}}]{Li2009a}%
  \BibitemOpen
  \bibfield  {author} {\bibinfo {author} {\bibfnamefont {X.}~\bibnamefont
  {Li}}, \bibinfo {author} {\bibfnamefont {W.}~\bibnamefont {Cai}}, \bibinfo
  {author} {\bibfnamefont {J.}~\bibnamefont {An}}, \bibinfo {author}
  {\bibfnamefont {S.}~\bibnamefont {Kim}}, \bibinfo {author} {\bibfnamefont
  {J.}~\bibnamefont {Nah}}, \bibinfo {author} {\bibfnamefont {D.}~\bibnamefont
  {Yang}}, \bibinfo {author} {\bibfnamefont {R.}~\bibnamefont {Piner}},
  \bibinfo {author} {\bibfnamefont {A.}~\bibnamefont {Velamakanni}}, \bibinfo
  {author} {\bibfnamefont {I.}~\bibnamefont {Jung}}, \bibinfo {author}
  {\bibfnamefont {E.}~\bibnamefont {Tutuc}}, \bibinfo {author} {\bibfnamefont
  {S.~K.}\ \bibnamefont {Banerjee}}, \bibinfo {author} {\bibfnamefont
  {L.}~\bibnamefont {Colombo}}, \ and\ \bibinfo {author} {\bibfnamefont
  {R.~S.}\ \bibnamefont {Ruoff}},\ }\href
  {http://dx.doi.org/10.1126/science.1171245} {\bibfield  {journal} {\bibinfo
  {journal} {Science},\ }\textbf {\bibinfo {volume} {324}},\ \bibinfo {pages}
  {1312} (\bibinfo {year} {2009})}\BibitemShut {NoStop}%
\bibitem [{\citenamefont {Gao}\ \emph {et~al.}(2010)\citenamefont {Gao},
  \citenamefont {Ren}, \citenamefont {Zhao}, \citenamefont {Ma}, \citenamefont
  {Chen},\ and\ \citenamefont {Cheng}}]{Gao2010a}%
  \BibitemOpen
  \bibfield  {author} {\bibinfo {author} {\bibfnamefont {L.}~\bibnamefont
  {Gao}}, \bibinfo {author} {\bibfnamefont {W.}~\bibnamefont {Ren}}, \bibinfo
  {author} {\bibfnamefont {J.}~\bibnamefont {Zhao}}, \bibinfo {author}
  {\bibfnamefont {L.-P.}\ \bibnamefont {Ma}}, \bibinfo {author} {\bibfnamefont
  {Z.}~\bibnamefont {Chen}}, \ and\ \bibinfo {author} {\bibfnamefont {H.-M.}\
  \bibnamefont {Cheng}},\ }\href {http://dx.doi.org/10.1063/1.3512865}
  {\bibfield  {journal} {\bibinfo  {journal} {Appl. Phys. Lett.},\ }\textbf
  {\bibinfo {volume} {97}},\ \bibinfo {pages} {183109} (\bibinfo {year}
  {2010})}\BibitemShut {NoStop}%
\bibitem [{\citenamefont {Celebi}\ \emph {et~al.}(2011)\citenamefont {Celebi},
  \citenamefont {Cole}, \citenamefont {Teo},\ and\ \citenamefont
  {Park}}]{Celebi2011}%
  \BibitemOpen
  \bibfield  {author} {\bibinfo {author} {\bibfnamefont {K.}~\bibnamefont
  {Celebi}}, \bibinfo {author} {\bibfnamefont {M.~T.}\ \bibnamefont {Cole}},
  \bibinfo {author} {\bibfnamefont {K.~B.~K.}\ \bibnamefont {Teo}}, \ and\
  \bibinfo {author} {\bibfnamefont {H.~G.}\ \bibnamefont {Park}},\ }\href
  {http://dx.doi.org/10.1149/2.005201esl} {\bibfield  {journal} {\bibinfo
  {journal} {Electrochem. Solid-State Lett.},\ }\textbf {\bibinfo {volume}
  {15}},\ \bibinfo {pages} {K1} (\bibinfo {year} {2011})}\BibitemShut {NoStop}%
\bibitem [{\citenamefont {Murdock}\ \emph {et~al.}(2013)\citenamefont
  {Murdock}, \citenamefont {Koos}, \citenamefont {Britton}, \citenamefont
  {Houben}, \citenamefont {Batten}, \citenamefont {Zhang}, \citenamefont
  {Wilkinson}, \citenamefont {Dunin-Borkowski}, \citenamefont {Lekka},\ and\
  \citenamefont {Grobert}}]{Murdock2013}%
  \BibitemOpen
  \bibfield  {author} {\bibinfo {author} {\bibfnamefont {A.~T.}\ \bibnamefont
  {Murdock}}, \bibinfo {author} {\bibfnamefont {A.}~\bibnamefont {Koos}},
  \bibinfo {author} {\bibfnamefont {T.~B.}\ \bibnamefont {Britton}}, \bibinfo
  {author} {\bibfnamefont {L.}~\bibnamefont {Houben}}, \bibinfo {author}
  {\bibfnamefont {T.}~\bibnamefont {Batten}}, \bibinfo {author} {\bibfnamefont
  {T.}~\bibnamefont {Zhang}}, \bibinfo {author} {\bibfnamefont {A.~J.}\
  \bibnamefont {Wilkinson}}, \bibinfo {author} {\bibfnamefont {R.~E.}\
  \bibnamefont {Dunin-Borkowski}}, \bibinfo {author} {\bibfnamefont {C.~E.}\
  \bibnamefont {Lekka}}, \ and\ \bibinfo {author} {\bibfnamefont
  {N.}~\bibnamefont {Grobert}},\ }\href {http://dx.doi.org/10.1021/nn3049297}
  {\bibfield  {journal} {\bibinfo  {journal} {{ACS} Nano},\ }\textbf {\bibinfo
  {volume} {7}},\ \bibinfo {pages} {1351} (\bibinfo {year} {2013})}\BibitemShut
  {NoStop}%
\bibitem [{\citenamefont {Tapaszto}\ \emph {et~al.}(2008)\citenamefont
  {Tapaszto}, \citenamefont {Dobrik}, \citenamefont {Lambin},\ and\
  \citenamefont {Biro}}]{Tapaszto2008}%
  \BibitemOpen
  \bibfield  {author} {\bibinfo {author} {\bibfnamefont {L.}~\bibnamefont
  {Tapaszto}}, \bibinfo {author} {\bibfnamefont {G.}~\bibnamefont {Dobrik}},
  \bibinfo {author} {\bibfnamefont {P.}~\bibnamefont {Lambin}}, \ and\ \bibinfo
  {author} {\bibfnamefont {L.~P.}\ \bibnamefont {Biro}},\ }\href
  {http://dx.doi.org/10.1038/nnano.2008.149} {\bibfield  {journal} {\bibinfo
  {journal} {Nature Nanotech.},\ }\textbf {\bibinfo {volume} {3}},\ \bibinfo
  {pages} {397} (\bibinfo {year} {2008})}\BibitemShut {NoStop}%
\bibitem [{\citenamefont {Fasoli}\ \emph {et~al.}(2009)\citenamefont {Fasoli},
  \citenamefont {Colli}, \citenamefont {Lombardo},\ and\ \citenamefont
  {Ferrari}}]{Fasoli2009}%
  \BibitemOpen
  \bibfield  {author} {\bibinfo {author} {\bibfnamefont {A.}~\bibnamefont
  {Fasoli}}, \bibinfo {author} {\bibfnamefont {A.}~\bibnamefont {Colli}},
  \bibinfo {author} {\bibfnamefont {A.}~\bibnamefont {Lombardo}}, \ and\
  \bibinfo {author} {\bibfnamefont {A.~C.}\ \bibnamefont {Ferrari}},\ }\href
  {http://dx.doi.org/10.1002/pssb.200982356} {\bibfield  {journal} {\bibinfo
  {journal} {Phys. Status Solidi B},\ }\textbf {\bibinfo {volume} {246}},\
  \bibinfo {pages} {2514} (\bibinfo {year} {2009})}\BibitemShut {NoStop}%
\bibitem [{\citenamefont {Kosynkin}\ \emph {et~al.}(2009)\citenamefont
  {Kosynkin}, \citenamefont {Higginbotham}, \citenamefont {Sinitskii},
  \citenamefont {Lomeda}, \citenamefont {Dimiev}, \citenamefont {Price},\ and\
  \citenamefont {Tour}}]{Kosynkin2009}%
  \BibitemOpen
  \bibfield  {author} {\bibinfo {author} {\bibfnamefont {D.~V.}\ \bibnamefont
  {Kosynkin}}, \bibinfo {author} {\bibfnamefont {A.~L.}\ \bibnamefont
  {Higginbotham}}, \bibinfo {author} {\bibfnamefont {A.}~\bibnamefont
  {Sinitskii}}, \bibinfo {author} {\bibfnamefont {J.~R.}\ \bibnamefont
  {Lomeda}}, \bibinfo {author} {\bibfnamefont {A.}~\bibnamefont {Dimiev}},
  \bibinfo {author} {\bibfnamefont {B.~K.}\ \bibnamefont {Price}}, \ and\
  \bibinfo {author} {\bibfnamefont {J.~M.}\ \bibnamefont {Tour}},\ }\href
  {http://dx.doi.org/10.1038/nature07872} {\bibfield  {journal} {\bibinfo
  {journal} {Nature},\ }\textbf {\bibinfo {volume} {458}},\ \bibinfo {pages}
  {872} (\bibinfo {year} {2009})}\BibitemShut {NoStop}%
\bibitem [{\citenamefont {Jiao}\ \emph {et~al.}(2009)\citenamefont {Jiao},
  \citenamefont {Zhang}, \citenamefont {Wang}, \citenamefont {Diankov},\ and\
  \citenamefont {Dai}}]{Jiao2009}%
  \BibitemOpen
  \bibfield  {author} {\bibinfo {author} {\bibfnamefont {L.}~\bibnamefont
  {Jiao}}, \bibinfo {author} {\bibfnamefont {L.}~\bibnamefont {Zhang}},
  \bibinfo {author} {\bibfnamefont {X.}~\bibnamefont {Wang}}, \bibinfo {author}
  {\bibfnamefont {G.}~\bibnamefont {Diankov}}, \ and\ \bibinfo {author}
  {\bibfnamefont {H.}~\bibnamefont {Dai}},\ }\href
  {http://dx.doi.org/10.1038/nature07919} {\bibfield  {journal} {\bibinfo
  {journal} {Nature},\ }\textbf {\bibinfo {volume} {458}},\ \bibinfo {pages}
  {877} (\bibinfo {year} {2009})}\BibitemShut {NoStop}%
\bibitem [{\citenamefont {Paiva}\ \emph {et~al.}(2010)\citenamefont {Paiva},
  \citenamefont {Xu}, \citenamefont {Fernanda~Proen\c{c}a}, \citenamefont
  {Novais}, \citenamefont {L{\ae}gsgaard},\ and\ \citenamefont
  {Besenbacher}}]{Paiva2010}%
  \BibitemOpen
  \bibfield  {author} {\bibinfo {author} {\bibfnamefont {M.~C.}\ \bibnamefont
  {Paiva}}, \bibinfo {author} {\bibfnamefont {W.}~\bibnamefont {Xu}}, \bibinfo
  {author} {\bibfnamefont {M.}~\bibnamefont {Fernanda~Proen\c{c}a}}, \bibinfo
  {author} {\bibfnamefont {R.~M.}\ \bibnamefont {Novais}}, \bibinfo {author}
  {\bibfnamefont {E.}~\bibnamefont {L{\ae}gsgaard}}, \ and\ \bibinfo {author}
  {\bibfnamefont {F.}~\bibnamefont {Besenbacher}},\ }\href
  {http://dx.doi.org/10.1021/nl100240n} {\bibfield  {journal} {\bibinfo
  {journal} {Nano Lett.},\ }\textbf {\bibinfo {volume} {10}},\ \bibinfo {pages}
  {1764} (\bibinfo {year} {2010})}\BibitemShut {NoStop}%
\bibitem [{\citenamefont {Melle-Franco}\ \emph {et~al.}(2004)\citenamefont
  {Melle-Franco}, \citenamefont {Marcaccio}, \citenamefont {Paolucci},
  \citenamefont {Paolucci}, \citenamefont {Georgakilas}, \citenamefont {Guldi},
  \citenamefont {Prato},\ and\ \citenamefont {Zerbetto}}]{Melle-Franco2004}%
  \BibitemOpen
  \bibfield  {author} {\bibinfo {author} {\bibfnamefont {M.}~\bibnamefont
  {Melle-Franco}}, \bibinfo {author} {\bibfnamefont {M.}~\bibnamefont
  {Marcaccio}}, \bibinfo {author} {\bibfnamefont {D.}~\bibnamefont {Paolucci}},
  \bibinfo {author} {\bibfnamefont {F.}~\bibnamefont {Paolucci}}, \bibinfo
  {author} {\bibfnamefont {V.}~\bibnamefont {Georgakilas}}, \bibinfo {author}
  {\bibfnamefont {D.~M.}\ \bibnamefont {Guldi}}, \bibinfo {author}
  {\bibfnamefont {M.}~\bibnamefont {Prato}}, \ and\ \bibinfo {author}
  {\bibfnamefont {F.}~\bibnamefont {Zerbetto}},\ }\href
  {http://dx.doi.org/10.1021/ja039918r} {\bibfield  {journal} {\bibinfo
  {journal} {J. Am. Chem. Soc.},\ }\textbf {\bibinfo {volume} {126}},\ \bibinfo
  {pages} {1646} (\bibinfo {year} {2004})}\BibitemShut {NoStop}%
\bibitem [{\citenamefont {Zhang}\ \emph {et~al.}(2013)\citenamefont {Zhang},
  \citenamefont {Yazyev}, \citenamefont {Feng}, \citenamefont {Xie},
  \citenamefont {Tao}, \citenamefont {Chen}, \citenamefont {Jiao},
  \citenamefont {Pedramrazi}, \citenamefont {Zettl}, \citenamefont {Louie},
  \citenamefont {Dai},\ and\ \citenamefont {Crommie}}]{Zhang2013}%
  \BibitemOpen
  \bibfield  {author} {\bibinfo {author} {\bibfnamefont {X.}~\bibnamefont
  {Zhang}}, \bibinfo {author} {\bibfnamefont {O.~V.}\ \bibnamefont {Yazyev}},
  \bibinfo {author} {\bibfnamefont {J.}~\bibnamefont {Feng}}, \bibinfo {author}
  {\bibfnamefont {L.}~\bibnamefont {Xie}}, \bibinfo {author} {\bibfnamefont
  {C.}~\bibnamefont {Tao}}, \bibinfo {author} {\bibfnamefont {Y.-C.}\
  \bibnamefont {Chen}}, \bibinfo {author} {\bibfnamefont {L.}~\bibnamefont
  {Jiao}}, \bibinfo {author} {\bibfnamefont {Z.}~\bibnamefont {Pedramrazi}},
  \bibinfo {author} {\bibfnamefont {A.}~\bibnamefont {Zettl}}, \bibinfo
  {author} {\bibfnamefont {S.~G.}\ \bibnamefont {Louie}}, \bibinfo {author}
  {\bibfnamefont {H.}~\bibnamefont {Dai}}, \ and\ \bibinfo {author}
  {\bibfnamefont {M.~F.}\ \bibnamefont {Crommie}},\ }\href
  {http://dx.doi.org/10.1021/nn303730v} {\bibfield  {journal} {\bibinfo
  {journal} {{ACS} Nano},\ }\textbf {\bibinfo {volume} {7}},\ \bibinfo {pages}
  {198} (\bibinfo {year} {2013})}\BibitemShut {NoStop}%
\bibitem [{\citenamefont {Xie}\ \emph {et~al.}(2010)\citenamefont {Xie},
  \citenamefont {Jiao},\ and\ \citenamefont {Dai}}]{Xie2010}%
  \BibitemOpen
  \bibfield  {author} {\bibinfo {author} {\bibfnamefont {L.}~\bibnamefont
  {Xie}}, \bibinfo {author} {\bibfnamefont {L.}~\bibnamefont {Jiao}}, \ and\
  \bibinfo {author} {\bibfnamefont {H.}~\bibnamefont {Dai}},\ }\href
  {http://dx.doi.org/10.1021/ja107071g} {\bibfield  {journal} {\bibinfo
  {journal} {J. Am. Chem. Soc.},\ }\textbf {\bibinfo {volume} {132}},\ \bibinfo
  {pages} {14751} (\bibinfo {year} {2010})}\BibitemShut {NoStop}%
\end{thebibliography}
\end{document}